\documentclass[twocolumn,prb,amsmath,amssymb,unsortedaddress,superscriptaddress,showpacs]{revtex4}

\usepackage{latexsym,epsfig,graphicx}% Include figure files
\usepackage{dcolumn}% Align table columns on decimal point
\usepackage{bm}% bold math
\usepackage{color} %coloring for comments
\usepackage{subfigure}
\begin{document}

\title{Transport in multi-terminal superconductor/ferromagnet
junctions having spin-dependent interfaces}

\author{Kuei Sun} \affiliation{Department of Physics, University of Cincinnati,
Cincinnati, Ohio 45221-0011, USA}
\author{Nayana Shah}\affiliation {Department of Physics, University of Cincinnati, Cincinnati,
Ohio 45221-0011, USA}
\author{Smitha Vishveshwara} \affiliation{Department of Physics, University of
Illinois at Urbana-Champaign, Urbana, Illinois 61801-3080, USA}
\date{February 22, 2013}


\begin{abstract}
We study electronic transport in junctions consisting of a
superconductor electrode and two ferromagnet (F) leads in which
crossed Andreev reflections (CAR) and elastic cotunnelings are
accommodated. We model the system using an extended
Blonder-Tinkham-Klapwijk (BTK) treatment with a key modification
that accounts for spin-dependent interfacial barriers (SDIB). We
compute current-voltage relations as a function of parameters
characterizing the SDIB, magnetization in the F leads, geometry of
the junction, and temperature. Our results reveal a rich range of
significantly altered physics due to a combination of interfering
spin-dependent scattering processes and population imbalance in
the ferromagnets, such as a significant enhancement in CAR current
and a sign change in the relative difference between resistance of
two cases having a antiparallel or parallel alignment of the
magnetization in the F leads, respectively. Our model accounts for
the surprising experimental findings of positive relative
resistance by M. Colci \emph{et al}. [Phys. Rev. B \textbf{85},
180512(R) (2012)] as well as previously measured negative relative
resistance results, both within sufficiently large parameter
regions.
\end{abstract}

\maketitle

\section{Introduction}\label{sec:intro}
Over the past decades, extensive theoretical and experimental
research has focused on electronic transport in
superconductor--normal metal (S-N) heterostructures, wherein
conducting electrons propagate through bulk materials and scatter
at interfaces~\cite{Blonder82,Beenakker97,Lambert98,Eschrig09}.
More recently, attention has turned to systems comprising an S
electrode in contact with multi-N terminals, in which a unique
scattering process known as the crossed Andreev reflection (CAR)
can
occur~\cite{Byers95,Takahashi99,Deutscher00,Recher01,Falci01,Melin02,Melin03,Dong03,Russo05,Morten06,Metalidis10}.
The CAR process involves an electron in an N terminal impinging
the S accompanied by a hole of opposite spin reflecting in another
N terminal separated within the superconducting coherence length,
and generating a Cooper pair in S . The non local and coherent
nature of CAR makes such a device a platform for exploring quantum
entanglement, providing potential application to quantum
computing~\cite{Burkard07}.

Another avenue of S-N junction studies which has recently gained
prominence is the presence of  ferromagnetic order in the normal
system. Transport physics can be significantly altered when the
normal system is a ferromagnet (F). For instance, the F bulk can
be created by spin imbalance, changing the populations involved in
the transport processes as well as spin-dependent interfaces that
modify both the proximity
effect~\cite{Buzdin05,Bergeret05,Cottet05,Linder09} and electron
scattering.~\cite{Fogelstrom00,Barash02,Kopu04,Faure06} In S-F
junctions, the interplay between ferromagnetic and superconducting
orders around the interface strongly affects Andreev
reflections~\cite{deJong95,Giroud98,Zutic00,Strijkers01,Kupferschmidt11}
(the direct process of an incoming electron reflecting as a hole
of opposite spin within the same terminal~\cite{Andreev64}), while
in S-F-S systems it can lead to Josephson $\pi$ junctions
sustaining negative critical
currents.~\cite{Ryazanov01,Kontos02,Robinson07,Kastening09}

\begin{figure}[t]
\centering
   \includegraphics[width=7.9cm]{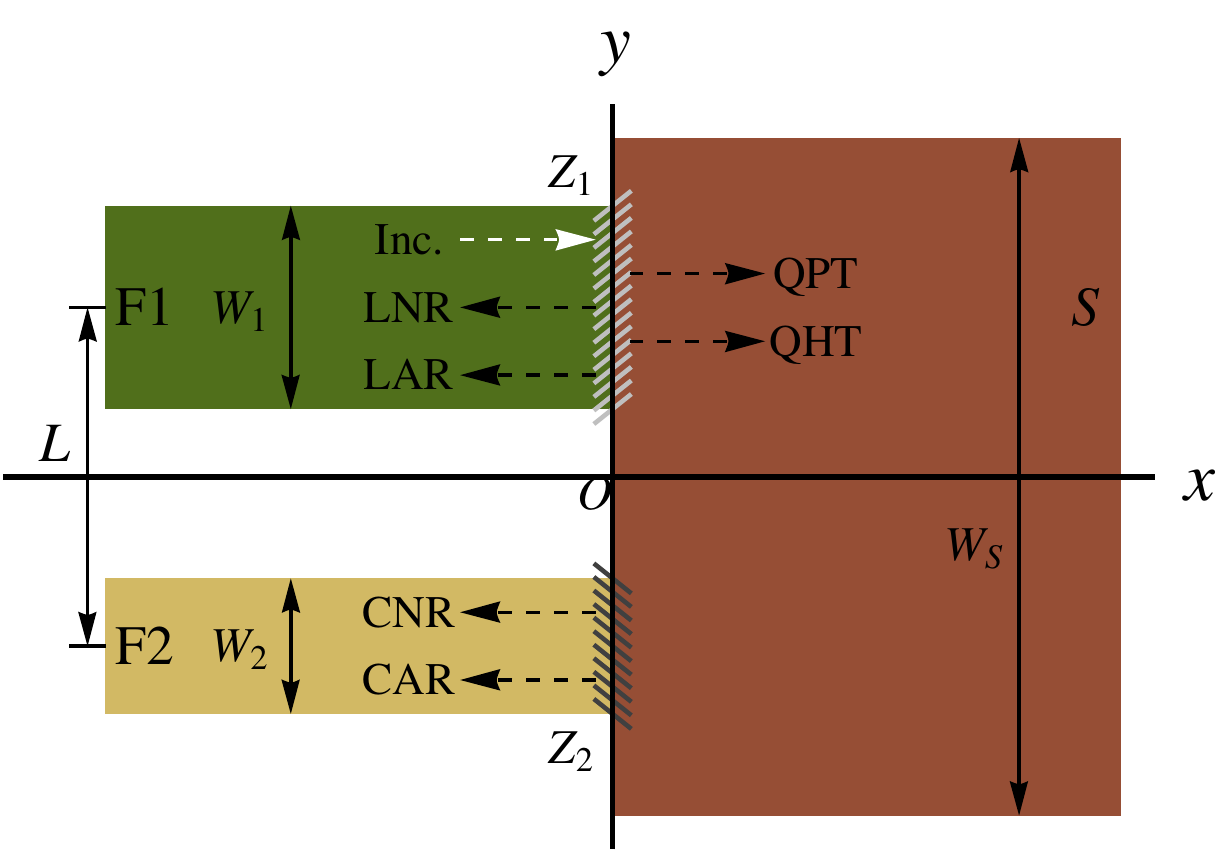}
  \caption{(Color online) Cartoon representation of an SFF
junction. The system has two ferro-magnetic leads F1 (F2),
illustrated as the shaded regions in the second (third) quadrant
having width $W_{1(2)}$ and separation $L$ between them, and a
superconducting electrode, $S$, illustrated as the shaded region
in the $x>0$ plane with width $W_S$.  The F1(2)-S interfaces
(twilled regions) are described by spin-dependent interface
parameters $Z_1$ and $Z_2$, respectively. Black dashed arrows
indicate six scattering processes experienced by an electron in F1
incident on the interface (Inc., white dashed arrow): local normal
reflection (LNR), local Andreev reflection (LAR), crossed normal
reflection (CNR, also referred to as elastic cotunneling), crossed
Andreev reflection (CAR), quasi particle transmission (QPT) and
quasi hole transmission (QHT).}
        \label{fig:f01}
\end{figure}

Recent experiments~\cite{Beckmann04,Luo09,Colci10,Colci12} have
been performed on S-FF-S devices in which two F bridges are both
laid across two S electrodes, with the separation between the two
F bridges being smaller than the superconducting coherence length
(thus capable of accommodating CAR). The results show that the
system can exhibit both positive~\cite{Colci10,Colci12} and
negative~\cite{Beckmann04,Luo09} relative resistances between two
different cases characterized by an anti parallel (AP) or parallel
(P) alignment of magnetization in the F leads, respectively. [The
relative resistance, denoted by $\delta R$, is defined as the
normalized value of the difference $R_{\rm{AP}}-R_{\rm{P}}$, where
$R_{\rm{AP(P)}}$ is the resistance across the junction with AP (P)
configuration. See Eq.~(\ref{eqn:DR}) for details.] Since coherent
transport between the two S electrodes is not revealed in the
experimental data, the S-FF-S device can be considered as a series
connection of two independent S-FF junctions (each of which, as
illustrated in Fig.~\ref{fig:f01}, carries coherent transport
between the two F leads). Therefore, one can study the transport
properties of the S-FF-S system by investigating the S-FF
junction. In such cases, the negative $\delta R$
($R_{\rm{P}}>R_{\rm{AP}}$) can be understood by an intuitive
picture that the suppression of CAR due to spin imbalance raises
the resistance of the P case, or can be explicitly explained by a
Blonder-Tinkham-Klapwijk (BTK) treatment~\cite{Blonder82} with
spin-independent interfacial barriers.~\cite{Yamashita03} However,
such a BTK model cannot explain the counter intuitive data of
positive $\delta R$, which suggests that a competitive effect
should be incorporated.

Motivated by these experiments, we provide one of the first
comprehensive theoretical studies of S-FF junctions having
spin-dependent interfacial barriers (SDIB). The presence of SDIB
at the junction between a superconductor and a single ferromagnet
has been shown to considerably alter transport
properties.~\cite{Fogelstrom00,Barash02,Kopu04,Faure06,deJong95,Giroud98,Zutic00,Strijkers01,Kupferschmidt11}
For our situation, the SDIB plays a prominent role in affecting
coherence and cross correlations between excitations in the two
ferromagnets. We explore the large parameter space of the system
and show that SDIB can give rise to a rich range of physics
depending on the choice of parameters. We provide a consistent
scenario explaining both competing experimental results described
above. In fact, a small subsection of the studies investigated
here has been directly employed to explain the counterintuitive
experimental results presented in Ref.~\onlinecite{Colci12}.  We
propose a microscopic picture describing the SDIB within an
extended BTK model and compute transport properties of the system
as a function of the derived SDIB parameters. We carefully discuss
the various scattering processes responsible for the difference in
behavior between the P and AP cases and show that interference
between these processes can play a significant role. We analyze
transport properties as a function of magnetization, the geometry
of the system, and temperature. We show that they reveal several
interesting features, such as, the sought-after differing behavior
for the P and AP cases, oscillations as a function of geometry,
and, under certain conditions, enhancement of crossed Andreev
reflection, which would be indicative of long-range entanglement.

Our studies thus not only reconcile the apparently conflicting
results of the various experiments; they also provide several
other aspects to explore within the same experimental settings.
Our extended BTK model also  applies to several other possible
experimental geometries, including  other multi terminal
superconducting hybrids that possess a two dimensional nature
characterizing the elongation and separation of the terminals,
such as multi layer~\cite{Aarts97,Gu02,Giazotto06} or multi
wire~\cite{Tian05,Altomare06} devices in the presence of the
spin-dependent interfacial scattering.

The paper is outlined as follows. In Sec.~\ref{sec:model} we model
the geometry of the system and present the corresponding
Hamiltonian describing six scattering processes. We also discuss a
possible microscopic mechanism resulting in an SDIB. In
Sec.~\ref{sec:BTK}, by applying the BTK treatment we derive the
current-voltage ($I$-$V$) relation as a function of exchange
energy, SDIB parameters, the geometry of the junction and
temperature. In Sec.~\ref{sec:results}, we illustrate data showing
the dependence of transport properties on several groups of
variables associated with  ferromagnetism, geometry of the system,
and temperature, respectively. We discuss the effects of SDIB
compared to the spin-independent case and other effects competing
or cooperating with SDIB. In Sec.~\ref{sec:conclusion} we
summarize our results and comment on the scope of our work.

\section{The Model}\label{sec:model}
To model our basic setup  shown in Fig.~\ref{fig:f01}, we consider
a two-dimensional junction consisting of an S electrode in contact
with two F leads. The S electrode of width $W_S$ is located in the
$x>0$ half plane, while the two F leads (F1 and F2), of widths
$W_1$ and $W_2$, respectively, are located in the $x<0$ half plane
and separated by $L$ in the ${{\bf{\hat y}}}$ direction. This
system is described by a Bogoliubov--de Gennes (BdG) Hamiltonian
\begin{eqnarray}
\left( {\begin{array}{*{20}{c}} {{H_0} - \sigma {\epsilon
^{{\rm{ex}}}}({\bf{r}})
+ V_\sigma ^{\rm{I}}({\bf{r}})}&{\Delta ({\bf{r}})}\\
{\Delta ({\bf{r}})}&{ - [{H_0} - \bar \sigma {\epsilon
^{{\rm{ex}}}}({\bf{r}}) + V_{\bar \sigma }^{\rm{I}}({\bf{r}})]}
\end{array}} \right). \label{eqn:HAM}
\end{eqnarray}
Here ${H_0} =  - \frac{{{\hbar ^2}}}{{2m}}{\nabla ^2} - {\epsilon
_F}$ is the free Hamiltonian with Fermi energy, $\epsilon_F$,
${\bf{r}} = (x,y)$ denotes the coordinates, $\sigma=\pm$ denotes
up (down) spin states, respectively, and $\bar \sigma \equiv
-\sigma$. The superconducting gap $\Delta$, exchange energy
$\epsilon ^{{\rm{ex}}}$, and interface potential $V_I$ are given
by
\begin{eqnarray}
\Delta ({\bf{r}}) &=& \Delta {\theta}(x){\theta _S}(y)\\
{\epsilon ^{{\rm{ex}}}}({\bf{r}}) &=& {\theta}( - x)
\sum\limits_{j = 1,2} {\epsilon _j^{{\rm{ex}}}{\theta _j}(y)}\\
V_\sigma ^{\rm{I}}({\bf{r}}) &=& \delta (x)\sum\limits_{j = 1,2}
{{Z_{j\sigma }}{\theta _j}(y)} ,
\end{eqnarray}
where $\epsilon _{j=1,2}^{{\rm{ex}}}$ are exchange energies in
F1(2), respectively, indicating the Zeeman splitting between up
and down spins induced by ferromagnetism. The parameter
 $Z_{j\sigma}$ describes the corresponding S-F
interfacial barrier, $\theta$ is the step function, and
\begin{eqnarray}
{\theta _{j}}(y) &\equiv& \theta \left(\frac{{{W_{j}}}}{2} -
\left|
{y \mp \frac{L}{2}} \right|\right),\\
{\theta _S}(y) &\equiv& \theta \left(\frac{{{W_S}}}{2} - \left| y
\right|\right).
\end{eqnarray}
Here the $\delta$ function and the step function are applied to
confine the physical quantities in the corresponding regions.

The SDIB parameters $Z_\sigma$ can be decomposed as
\begin{eqnarray}
{Z_\sigma } = Z_0 \pm Z_s,\label{eqn:Zsig}
\end{eqnarray}
where the spin-independent component $Z_0$ represents effects of
an oxide layer or the local disorder at the
interface,~\cite{Blonder82} while the spin-dependent one $Z_s$ can
be induced by ferromagnetism. The sign depends on whether a
particular spin component $\sigma$ is the majority ($+$) or
minority ($-$) carrier. As one of the prevalent causes for an
SDIB, we propose a semiclassical picture to derive $Z_s$ as a
function of the average deviation of exchange energy compared with
a no-barrier case, $\langle \delta {{\epsilon ^{{\rm{ex}}}}}
\rangle$, over a microscopic length, $\xi$, associated with the
interfacial properties:
\begin{eqnarray}
{Z_s } =  \tan \left( {\frac{{  {\langle {{\delta \epsilon
^{{\rm{ex}}}}} \rangle} \xi }}{{\hbar {v_F}}}}
\right). \label{eqn:Zp}
\end{eqnarray}
This form can be derived by considering the phase accumulation of
an electron in a microscopic model passing through a finite layer
having spin-dependent potentials and relating it to the phase
shift in an effective model described by the BTK $\delta$ function
barrier~\cite{Tokuyasu88} (see Appendix \ref{sec:semiclassical}).
Equations (\ref{eqn:Zsig}) and (\ref{eqn:Zp}) show that for a
purely magnetic interface ($Z_0=0$) one has ${Z_ \downarrow } = -
{Z_ \uparrow }$ with the magnitude sensitive to the microscopic
configuration.

Now we turn to six different scattering processes through the
interface, which describe the transport physics in the system. As
illustrated in Fig.~\ref{fig:f01}, given an electron of energy
lower than the superconducting gap injected from F1, there are two
local scattering processes at the F1-S interface, (1) the direct
backscattering, or local normal reflection (LNR), and (2) local
Andreev reflection (LAR), as well as two crossed scattering
processes mediated by the superconducting order at the F2-S
interface, (3) electron backscattering called elastic
co-tunneling, and (4) crossed Andreev reflection (CAR). For
convenience in comparison, we refer to the elastic co-tunneling as
crossed normal reflection (CNR) in what follows. If the energy is
higher than the superconducting gap, we have two more processes:
(5) quasi particle transmissions (QPT) and (6) quasi hole
transmissions (QHT) into the S region. In Sec.~\ref{sec:BTK} we
apply the BTK treatment to compute scattering amplitudes subject
to the incoming state as well as the SDIB and hence obtain a net
charge current carried by these scattering processes.

\section{BTK treatment}\label{sec:BTK}

The BTK treatment is to consider charge transport as a net effect
of reflections and transmissions of electrons or holes at the N-S
(F-S) interface. The scattering amplitudes of reflections and
transmissions are obtained by solving the BdG equation with the
interface potential. In the presence of voltage drop across the
interface, an induced current is computed by summing the
probability current contributed by each scattering process,
weighted by the Fermi distribution. The calculations are aimed at
a current-voltage relation as a function of physical variables of
interests, such as the interface parameters, exchange energy,
geometry of the system, and temperature.

We start with the Hamiltonian of Eq.~(\ref{eqn:HAM}) for
considering incoming and outgoing waves incorporating the six
scattering processes (discussed in Sec.~\ref{sec:model}). As
prescribed by the BTK treatment, we solve for the forms of the
incoming and outgoing waves by boundary condition matching. Our
theoretical set-up requires careful accounting of the
spin-species, the multiple channels and the SDIB; we thus provide
a detailed outline of the procedure below.

We consider an incoming wave of energy $E$ from the F side
($x<0$). The wave function is of the form of a plane wave in the
${\bf{\hat x}}$ direction multiplied by a bound wave in the
${\bf{\hat y}}$ direction. We use indexes $\tau,\sigma$ indicating
a particle ($\tau=+$) with spin $\sigma$ or a hole ($\tau=-$) with
spin $\bar \sigma$, the channel number $n$ labeling the bound
states in the ${\bf{\hat y}}$ direction, and an index $j$ denoting
the wave in F$j$ ($j=1,2$) regions (see Fig.~\ref{fig:f01}). The
incoming wave is written as
\begin{eqnarray}
\Psi _{\tau \sigma j n}^{{\rm{in}}} = \left[ {\left(
{\begin{array}{*{20}{c}}
{{\delta _{\tau  + }}}\\
0
\end{array}} \right){e^{ixp_{\sigma ,n}^{+,j}}} + \left( {\begin{array}{*{20}{c}}
0\\
{{\delta _{\tau  - }}}
\end{array}} \right){e^{ - ixp_{ \bar \sigma ,n}^{-,j}}}}
\right]\Phi _n^j(y),\label{eqn:WFin}\nonumber\\
\end{eqnarray}
where $\delta$ is the $\delta$ function. The wave vectors $p$ and
the $y$-component wave function $\Phi$ are given below in
Eqs.~(\ref{eqn:WV12}) and (\ref{eqn:WFy12}), respectively. The
sign in front of of the wave vector is chosen to match the
direction of the group velocity.

The outgoing wave is represented as a linear combination of
degenerate scattering modes with energies and group velocities
corresponding to the incoming wave. The wave functions ${\Psi
_{j/\rm{S}}}({\bf{r}})$ in F$j$/S regions are separately given as
\begin{eqnarray}
\Psi _j ^{{\rm{out}}} &=& \sum\limits_{l = 1}^{M_a^j} {a_l^j\left(
{\begin{array}{*{20}{c}}
0\\
1
\end{array}} \right){e^{ixp_{ \bar \sigma ,l}^{-,j}}}\Phi _l^j(y)}\nonumber\\
&{}& + \sum\limits_{l = 1}^{M_b^j} {b_l^j\left(
{\begin{array}{*{20}{c}}
1\\
0
\end{array}} \right){e^{ - ixp_{\sigma ,l}^{+,j }}}\Phi _l^j(y)} ,\label{eqn:WF12}
\\
\Psi _{\rm{S}}^{{\rm{out}}} &=& \sum\limits_{l = 1}^{{M_{\rm{S}}}}
{\left[ {{c_l}\left( {\begin{array}{*{20}{c}}
{{u}}\\
{{v}}
\end{array}} \right){e^{ixk_l^ + }} + {d_l}\left( {\begin{array}{*{20}{c}}
{{v}}\\
{{u}}
\end{array}} \right){e^{ - ixk_l^ - }}} \right]\Phi _l^{\rm{S}}(y)} .\label{eqn:WFS}\nonumber\\
\end{eqnarray}
Here $a^j_l$ are the amplitudes for LAR (CAR) of channel $l$ if
$j$ is the same as (different from) the incoming wave, and
similarly $b^j_l$ represent the LNR (CNR) processes. The
amplitudes $c_l$ ($d_l$) correspond to quasi-particle (quasi hole)
transmissions in channel $l$. These amplitudes of the outgoing
waves are also functions of the indices $\{\tau,\sigma,j,n\}$ of
the incoming wave, which have been dropped here for convenience.
The quasi particle basis $u$ and $v$ satisfies $u^2 = 1 - v^2 =
\frac{1}{2}( {1 + \sqrt {1 - {{{\Delta ^2}}}/{{{E^2}}}} } )$. The
wave vectors of the $x$ component of the wave functions in channel
$l$ are given by
\begin{eqnarray}
p_{\sigma, l}^ {\tau, j}  &=& \sqrt {1 + \tau E + \sigma
{\epsilon _j^{{\rm{ex}}}} - ({l \pi}/{W_{j}})^2},\label{eqn:WV12}\\
k_l^ \tau  &=& \sqrt {1 + \tau \sqrt {{E^2} - {\Delta ^2}}  - ({l
\pi}/{W_{\rm{S}}})^2}.\label{eqn:WVS}
\end{eqnarray}
From here on, we take the Fermi energy and inverse of the Fermi
wave vector as energy and length units
($\epsilon_{\rm{F}}=k_{\rm{F}}^{-1}=1$), respectively. In
Eq.~(\ref{eqn:WF12}), the upper bounds of the summations,
$M_{a/b}^j$, are given by the highest current-carrying mode, above
which a mode has a purely imaginary wave vector and hence carries
no current. In Eq.~(\ref{eqn:WFS}), the wave vector can never be
purely imaginary if $E<\Delta$. In such case, we choose
$M_{\rm{S}}$ large enough to guarantee the convergence in
numerical calculations.~\cite{Yamashita03} The $y$ components of
the wave functions in Eq.~(\ref{eqn:WV12}) and (\ref{eqn:WVS}) are
given by
\begin{eqnarray}
{\Phi ^{j}_l} &=& \sqrt {\frac{2}{{{W_{j}}}}} \sin \left[l\pi
\left( {\frac{{y \mp L/2}}{{{W_{j}}}} + \frac{1}{2}}
\right)\right]{\theta _{j}}(y),\label{eqn:WFy12}\\
{\Phi ^{\rm{S}}_l} &=& \sqrt {\frac{2}{{{W_S}}}} \sin\left[ l\pi
\left( {\frac{y}{{{W_S}}} + \frac{1}{2}}\right) \right]{\theta
_S}(y).\label{eqn:WFyS}
\end{eqnarray}

To solve for the amplitudes, we match the incoming and outgoing
waves by imposing the boundary conditions at the interface
($x=0$):
\begin{eqnarray}
&{}&\Psi_{{\tau \sigma jn}}^{{\rm{in}}}(0,y) + \sum\limits_{i =
1}^2
{\Psi _i^{{\rm{out}}}(0,y)}  = \Psi _{\rm{S}}^{{\rm{out}}}(0,y), \label{eqn:BD1}\\
&{}&\left[ {{\partial _x} + \left( {\begin{array}{*{20}{c}}
{{Z_{j\sigma }}}&0\\
0&{{Z_{j \bar \sigma }}}
\end{array}} \right)} \right]\Psi _{\tau \sigma
nj}^{{\rm{in}}}(0,y)\nonumber\\
&{}& + \sum\limits_{i = 1}^2 {\left[ {{\partial _x} + \left(
{\begin{array}{*{20}{c}}
{{Z_{i\sigma }}}&0\\
0&{{Z_{i \bar \sigma }}}
\end{array}} \right)} \right]\Psi _j^{{\rm{out}}}(0,y)}  = {\partial _x}\Psi _{\rm{S}}^{{\rm{out}}}(0,y)\label{eqn:BD2}.\nonumber\\
\end{eqnarray}
Here Eq.~(\ref{eqn:BD1}) is the continuity equation, while
Eq.~(\ref{eqn:BD2}) is obtained from integrating the BdG equation
through the interface ($\int_{{0^ - }}^{{0^ + }} {dx}$). For
$Z_{i\sigma }=Z_{i \bar \sigma }$, Eq.~(\ref{eqn:BD2}) reduces to
the boundary condition for the case of a spin-independent
interface discussed in Ref.~\onlinecite{Yamashita03}. We
substitute the wave functions in
Eqs.~(\ref{eqn:WFin})--(\ref{eqn:WFS}) into Eq.~(\ref{eqn:BD1})
and project it onto channel $m$ in the S region,
\begin{eqnarray}
&&\left( {\begin{array}{*{20}{c}}
{{\delta _{\tau  + }}}\\
{{\delta _{\tau  - }}}
\end{array}} \right)\Lambda _{nm}^j \nonumber\\
&&+ \sum\limits_{i = 1}^2 {\left[ {\left( {\begin{array}{*{20}{c}}
0\\
1
\end{array}} \right)\sum\limits_{l = 1}^{M_a^i} {\Lambda _{lm}^ia_l^i + \left( {\begin{array}{*{20}{c}}
1\\
0
\end{array}} \right)\sum\limits_{l = 1}^{M_b^i} {\Lambda _{lm}^ib_l^i} } } \right]}
\nonumber\\
&&= \left( {\begin{array}{*{20}{c}}
{{u}}\\
{{v}}
\end{array}} \right){c_m} + \left( {\begin{array}{*{20}{c}}
{{v}}\\
{{u}}
\end{array}} \right){d_m}
,\label{eqn:AMS}
\end{eqnarray}
where
\begin{eqnarray}
\Lambda _{lm}^{j} = \int {dy{\Phi ^{j}_l}(y){\Phi
^{\rm{S}}_m}(y)}.\label{eqn:LAMBDA}
\end{eqnarray}
Similarly, we substitute the wave functions into
Eq.~(\ref{eqn:BD2}) and project it onto the channel $m$ in the
F$i$ region,
\begin{eqnarray}
&&\left( {\begin{array}{*{20}{c}}
{p_{\sigma ,n}^{ + ,j} - i{Z_{j\sigma }}}&0\\
0&{ - p_{ \bar \sigma ,n}^{ - ,j} - i{Z_{j \bar \sigma }}}
\end{array}} \right)\left( {\begin{array}{*{20}{c}}
{{\delta _{\tau  + }}}\\
{{\delta _{\tau  - }}}
\end{array}} \right){\delta _{ij}}{\delta _{mn}}\nonumber\\
&& +  \left( {\begin{array}{*{20}{c}}
{ - p_{\sigma ,m}^{ + ,i} - i{Z_{i\sigma }}}&0\\
0&{p_{ \bar \sigma ,m}^{ - ,i} - i{Z_{i \bar \sigma }}}
\end{array}} \right)\nonumber\\
&& \times \left[ {\left( {\begin{array}{*{20}{c}}
0\\
1
\end{array}} \right)a_m^i + \left( {\begin{array}{*{20}{c}}
1\\
0
\end{array}} \right)b_m^i} \right]\nonumber\\
&& = \sum\limits_{l = 1}^{{M_{\rm{S}}}} {\Lambda _{ml}^i\left[
{k_l^ + \left( {\begin{array}{*{20}{c}}
u\\
v
\end{array}} \right){c_l} - k_l^ - \left( {\begin{array}{*{20}{c}}
v\\
u
\end{array}} \right){d_l}} \right]}. \label{eqn:AM12}
\end{eqnarray}
Using Eq.~(\ref{eqn:AMS}) to replace $c_l$ and $d_l$ in
Eq.~(\ref{eqn:AM12}), we rewrite the right-hand side of
Eq.~(\ref{eqn:AM12}) as
\begin{eqnarray} &&
\Omega _{mn}^{ij}\left( {\begin{array}{*{20}{c}}
{{\delta _{\tau  + }}}\\
{{\delta _{\tau  - }}}
\end{array}} \right)\nonumber\\
&& + \sum\limits_{k = 1}^2 {\left[ {\sum\limits_{l = 1}^{M_a^k}
{\Omega _{ml}^{ik}\left( {\begin{array}{*{20}{c}}
0\\
1
\end{array}} \right)a_l^k + \sum\limits_{l = 1}^{M_b^k} {\Omega _{ml}^{ik}\left( {\begin{array}{*{20}{c}}
1\\
0
\end{array}} \right)b_l^k} } }
\right]},
\end{eqnarray}
where $\Omega _{mn}^{ij}$ is a $2 \times 2$ matrix defined as
\begin{eqnarray}
&& \Omega _{mn}^{ij} \equiv \left( {\begin{array}{*{20}{c}}
u&v\\
v&u
\end{array}} \right)\left[ {\sum\limits_{l = 1}^{{M_{\rm{S}}}} {\Lambda _{ml}^i\Lambda _{nl}^j\left( {\begin{array}{*{20}{c}}
{k_l^ + }&0\\
0&{ - k_l^ - }
\end{array}} \right)} } \right]\nonumber\\
&& \times {\left( {\begin{array}{*{20}{c}}
u&v\\
v&u
\end{array}} \right)^{ - 1}}.\label{eqn:AM12b}
\end{eqnarray}
Combining Eqs.~(\ref{eqn:AM12})--(\ref{eqn:AM12b}) and letting $i$
run from $1$ to $2$ as well as $m$ run through all
current-carrying channels, we get a set of linear equations for
obtaining the amplitudes $a^i_m$ and $b^i_m$.

Next we turn to compute the probability currents, which in turn
yield the current-voltage relation. The probability currents for
wave functions of a Nambu form, ${\left( {\begin{array}{*{20}{c}}
{{\psi _1}}&{{\psi _2}}
\end{array}} \right)^{\rm{T}}}$, are defined as
\begin{eqnarray}
{\bf{J}} = \frac{\hbar }{m}{\mathop{\rm Im}\nolimits} (\psi
_1^*\nabla {\psi _1} - \psi _2^*\nabla {\psi _2}),
\end{eqnarray}
where the $\psi_1$ term represents the contribution from electrons
and the $\psi_2$ term does that of holes. The wave functions in
Eqs.~(\ref{eqn:WFin})--(\ref{eqn:WFS}) carry no current in the
${{\bf{\hat y}}}$ direction. The outgoing probability currents in
the ${{\bf{\hat x}}}$ direction at the interface ($x=0$) in the
ferromagnet and superconductor regions [$\tilde J(E)$ and
${{\tilde J}^{{\rm{S}}}}(E)$, respectively] are given by
\begin{eqnarray}
\tilde J_{\tau \sigma jn}^{\tau '\sigma 'j'n'} &=& \frac{\hbar
}{m}\bigg[{\delta _{\tau ' - }}{\left| {a_{n'}^{j'}}
\right|^2}{\mathop{\rm Re}\nolimits} (p_{\sigma ',n'}^{ -
,j'})\nonumber\\
&{}& - {\delta _{\tau ' + }}{\left| {b_{n'}^{j'}}
\right|^2}{\mathop{\rm Re}\nolimits} (p_{\sigma ',n'}^{ +
,j'})\bigg]/J_{\tau \sigma jn}^{{\rm{in}}},\label{eqn:PCF}\\
\tilde J_{\tau \sigma jn}^{{\rm{S;}}\tau 'n'} &=& \frac{\hbar
}{m}\left[{\delta _{\tau ' + }}{\left| {{c_{n'}}}
\right|^2}{\mathop{\rm Re}\nolimits} (k_{n'}^ + ) + {\delta _{\tau
' - }}{\left| {{d_{n'}}} \right|^2}{\mathop{\rm Re}\nolimits}
(k_{n'}^ - )\right]\nonumber\\
&{}&\times \left({\left| u \right|^2} - {\left| v
\right|^2}\right)/J_{\tau \sigma jn}^{{\rm{in}}},\label{eqn:PCS}
\end{eqnarray}
normalized by the incoming current,
\begin{eqnarray}
J_{\tau \sigma jn}^{{\rm{in}}} = \frac{\hbar }{m}\left({\delta
_{\tau + }}p_{\sigma ,n}^{ + ,j} + {\delta _{\tau  - }}p_{\bar
\sigma ,n}^{ - ,j}\right).\label{eqn:PCIn}
\end{eqnarray}
Here the subscripts (superscripts) denote the corresponding
incoming (outgoing) state. In Eq.~(\ref{eqn:PCF}), $\{ \tau
',\sigma '\} = \pm \{ \tau ,\sigma \}$ represents normal or
Andreev reflections, respectively, while $j' = j$ or $\bar j$ (the
counterpart of $j$) denotes local or crossed processes,
respectively.

We assume the bias voltage $V$ to be the same across both F leads
and the S electrode, as set up in the experiment of
Ref.~\onlinecite{Colci12}. We remark that our method can also be
applied to the case of different voltages on different leads. Such
a scenario could enable one to probe other physical properties,
such as current correlation~\cite{Taddei02} and long-range
entanglement, in the system.

Following the standard BTK formalism,~\cite{Blonder82,Yamashita03}
we obtain the charge current, $I$, carried by incoming particles
or holes ($\tau=\pm$) with $\sigma$ spin in channel $n$ in lead
F$j$ (Appendix \ref{sec:I-V}),
\begin{eqnarray}
&{}&{I_{\tau \sigma jn}}(V) =\tau e \int\limits_0^\infty  {dE}
\nonumber\\
&{}&\times \big \{ (1 - \sum\limits_{j'} {\sum\limits_{n'} \left|
{\tilde J_{\tau \sigma jn}^{\tau \sigma j'n'}} \right| } )[f_0(E -
\tau eV) -
f_0(E)]\nonumber\\
&{}&+ \sum\limits_{j'} {\sum\limits_{n'} \left|{\tilde J_{\tau
\sigma jn}^{{\rm{ }}\bar \tau \bar \sigma j'n'}}\right| } [f_0(E)
- f_0(E + \tau eV)]\big \},\label{eqn:Ipar}
\end{eqnarray}
up to a constant associated with density of states, Fermi velocity
and an effective cross-sectional area~\cite{Yamashita03}. Here
${f_0}(E) = {[\exp (E/k_B T) + 1]^{ - 1}}$ is the Fermi
distribution function at temperature $T$, and $\bar \tau = -\tau$.
The total charge current is obtained by summing over all
contributions from the incoming waves,
\begin{eqnarray}
I(V) = \sum\limits_{\tau  =  \pm } {\sum\limits_{\sigma  =  \pm }
{\sum\limits_{j = 1,2} {\sum\limits_n {{I_{\tau \sigma jn}}} } }
}.\label{eqn:Itot}
\end{eqnarray}
Notice that $I$ is also a function of $\Delta$,
$\epsilon_{\rm{ex}}$, $T$, $Z_{j\sigma}$ and the geometry of the
system (characterized by $L$ and $W_{1/2/\rm{S}}$ here). Equipped
with this form for the current across the interface, we are now in a position
to explore it under various conditions.

\section{Results}\label{sec:results}

In this section we present major differences in transport
properties resulting from SDIB, compared to the previous BTK
studies.~\cite{Yamashita03} The key physical quantity showing such
differences is the normalized relative resistance, $\delta R$,
defined as resistance difference between two cases of the device
that have parallel (P) or anti parallel (AP) magnetization
alignment of the two F leads,
\begin{eqnarray}
\delta R=\frac{R_{\rm{AP}}-R_{\rm{P}}}{(R_{\rm{AP}}+R_{\rm{P}})/2}
=\frac{I_{\rm{P}}-I_{\rm{AP}}}{(I_{\rm{P}}+I_{\rm{AP}})/2},\label{eqn:DR}
\end{eqnarray}
where $I_{\rm{P(AP)}}$ is the total current obtained from
Eq.~(\ref{eqn:Itot}) for the corresponding cases and
$R_{\rm{P(AP)}}=V/I_{\rm{P(AP)}}$. The P (AP) case is
characterized by the same (opposite) signs of the exchange
energies in the two F leads, which leads to their majority spin
species being the same (opposite). Previous calculations in
Ref.~\onlinecite{Yamashita03} shows that $\delta R$ is always
non-positive, no matter how the barrier strength, exchange energy,
geometry of the device, and temperature vary. This result is as
expected of singlet-paired superconductors and in and of itself
would indicate that the AP case always carries more current than
the P case given the same bias voltage, in conflict with the
experimental results in Ref.~\onlinecite{Colci12}. Here we show
the exact manner in which the SDIB would alter this expected
trend.

In the following calculations, we take the majority spin in F1 to
be $\uparrow$, $\epsilon^{\rm{ex}}_2=\pm \epsilon^{\rm{ex}}_1$ for
the P (AP) case, and $eV=10^{-2}\Delta({\rm{at\ }}0T)=2 \times
10^{-5} \epsilon_F$ (corresponding to a Pippard superconducting
coherence length~\cite{Tinkham96,Leggett06} $\sim 500k_F^{-1}$ in
our case), all of which are typical values in experiments. We
first consider zero temperature where the system is in the subgap
regime and no quasi-particle transmission occurs. We analyze the
effect of two factors, ferromagnetism as well as geometry, on
which $\delta R$ shows strong dependence and discuss how SDIB are
relevant to the results. Then we study the finite-temperature case
where $\Delta$ varies in temperature and quasi particle
transmission contributes to the conductivity.

\begin{figure}[t]
\centering
   \includegraphics[width=6.5cm]{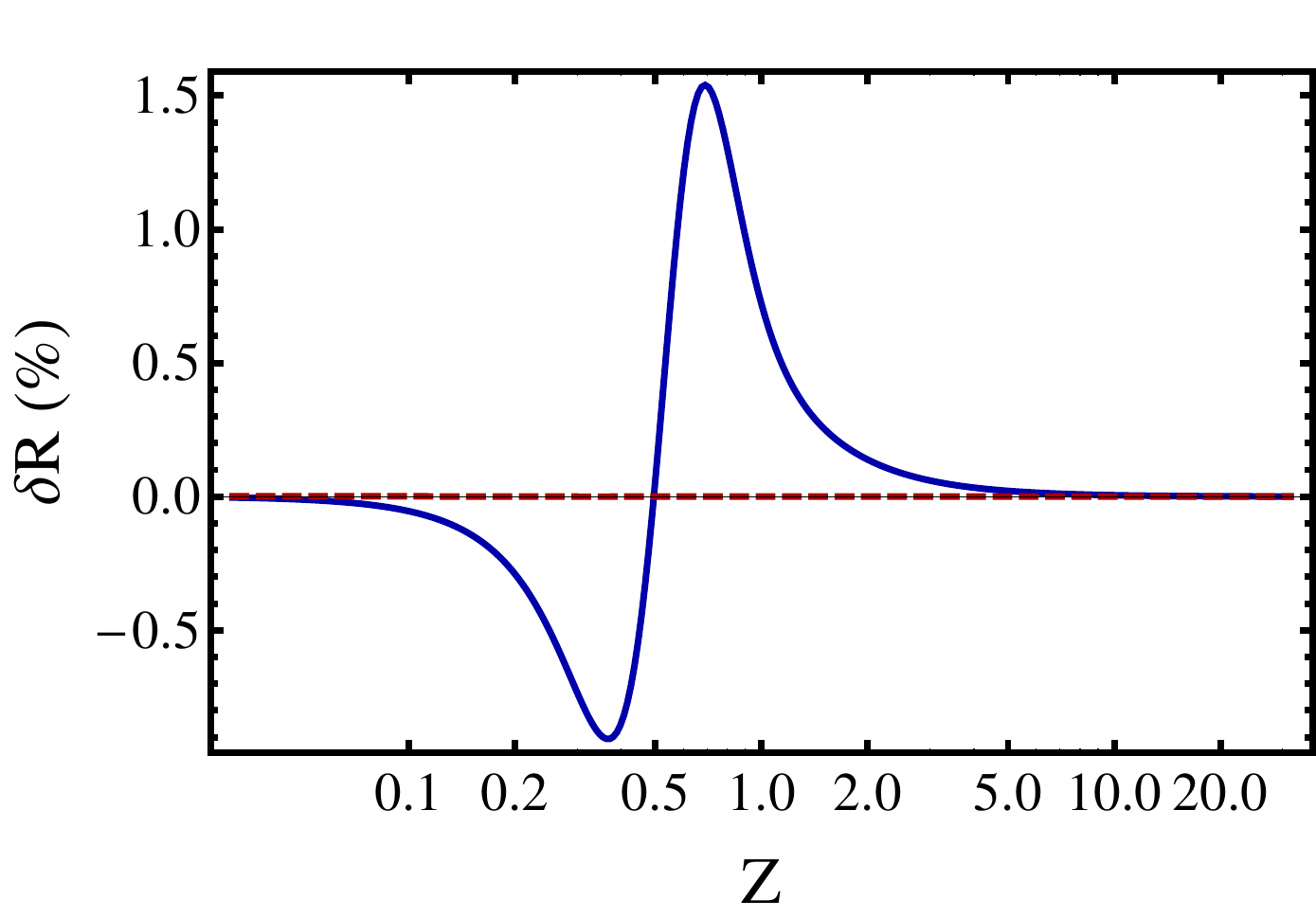}
  \caption{(Color online) Relative resistance $\delta R$
  (defined as $(R_{\rm{AP}}-R_{\rm{P}})/[(R_{\rm{AP}}+R_{\rm{P}})/2]$)
  vs interface parameters $Z$.
  The solid curve is for a purely spin-dependent barrier ($Z_0=0$, $Z_s=Z$)
  while the dashed curve is for a spin-independent one ($Z_0=Z$,
  $Z_s=0$). Data presented are for
  the system parameters $W_{1}=W_{2}=L=10k_F^{-1}$, $W_S=100k_F^{-1}$,
  $\epsilon^{\rm{ex}}=0.01\epsilon_F$, $\Delta=0.002\epsilon_F$, and $T=0$.
  }
        \label{fig:f02}
\end{figure}

\begin{figure}[t]
\centering
   \includegraphics[width=7.2cm]{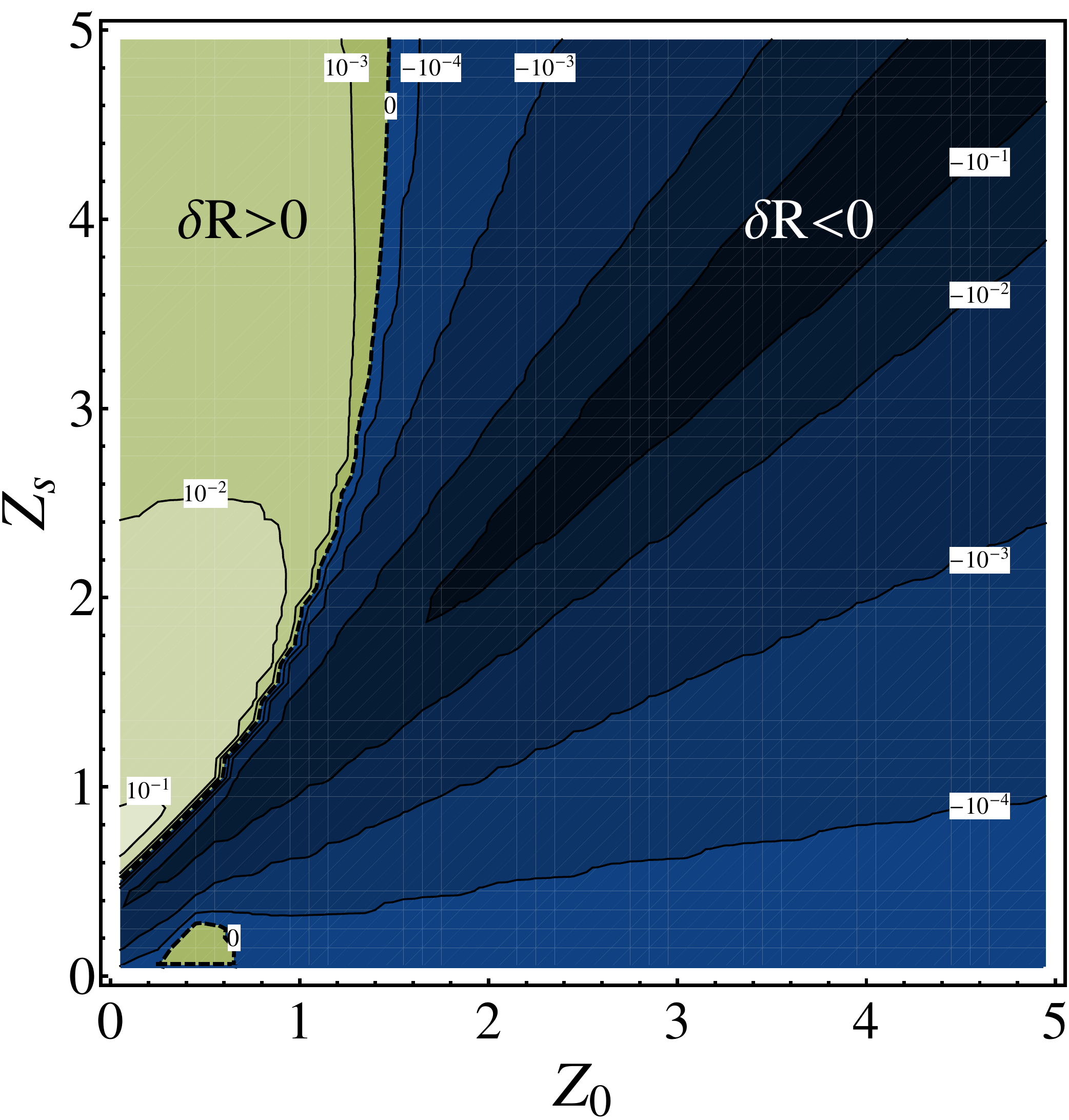}
  \caption{(Color online) Contour plot of $\delta R$ in the plane of $Z_0$ and $Z_s$.
  The positive (lighter green) and negative (darker blue) regions
  are separated by a dashed contour indicating $\delta R=0$.
  A weakly positive $\delta R$
  region appears close to the line of $Z_s=0$, for which $\delta R$
  is always negative.}
        \label{fig:f03}
\end{figure}

\begin{figure}[t]
\centering
   \includegraphics[width=6.5cm]{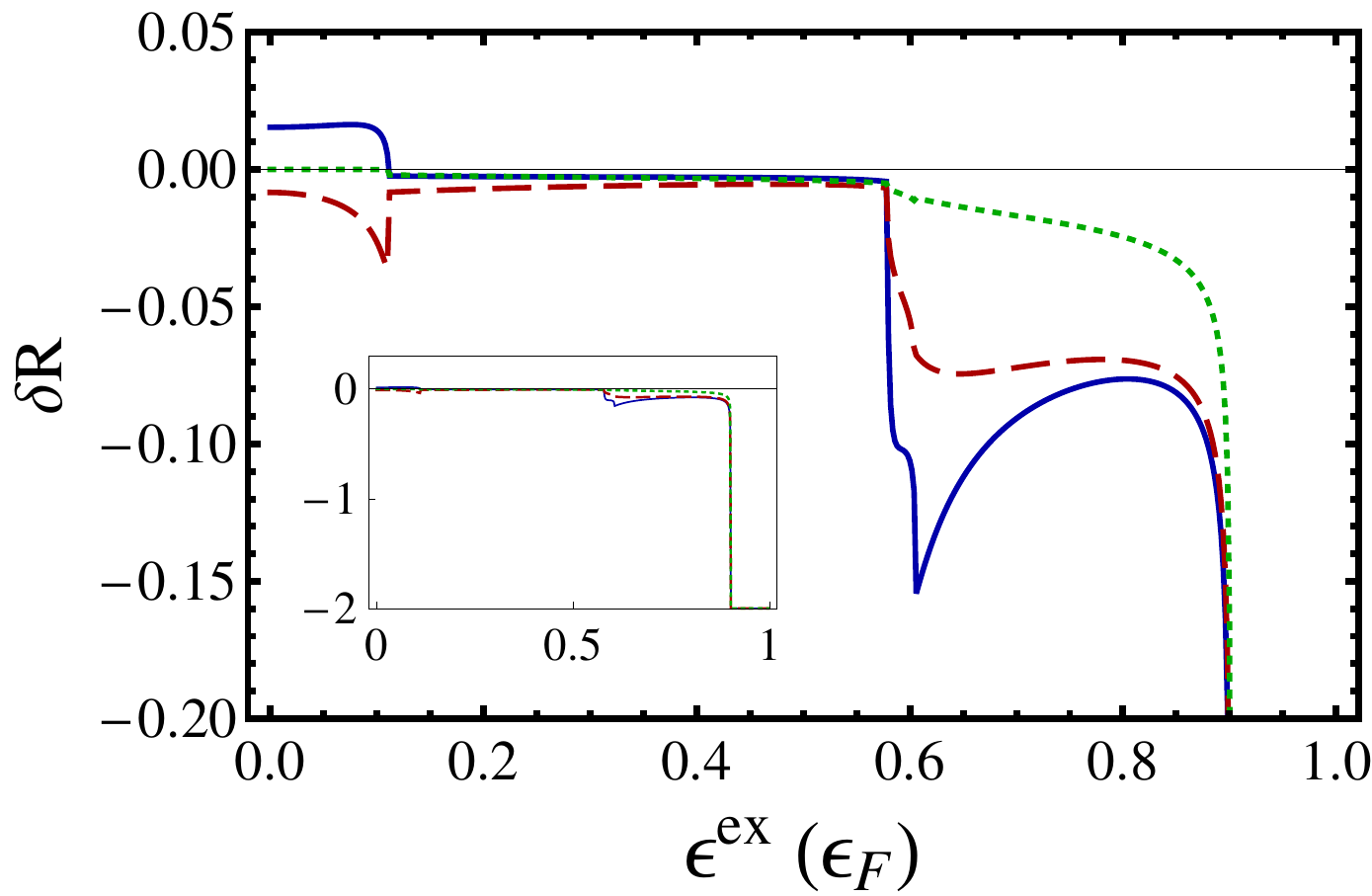}
  \caption{(Color online) Relative resistance $\delta R$
  vs exchange energy $\epsilon^{\rm{ex}}$ at the interface
  parameters $\{Z_0,Z_s\}=\{0,0.7\}$ (solid curve),
  $\{0,0.4\}$ (dashed), and $\{0.7,0\}$ (dotted). Inset: A
  zoom-out that shows $\delta R$ dropping to $-2$ at large
  imbalance.
  }
        \label{fig:f04}
\end{figure}

\begin{figure*}[t]
\centering
   \includegraphics[width=17cm]{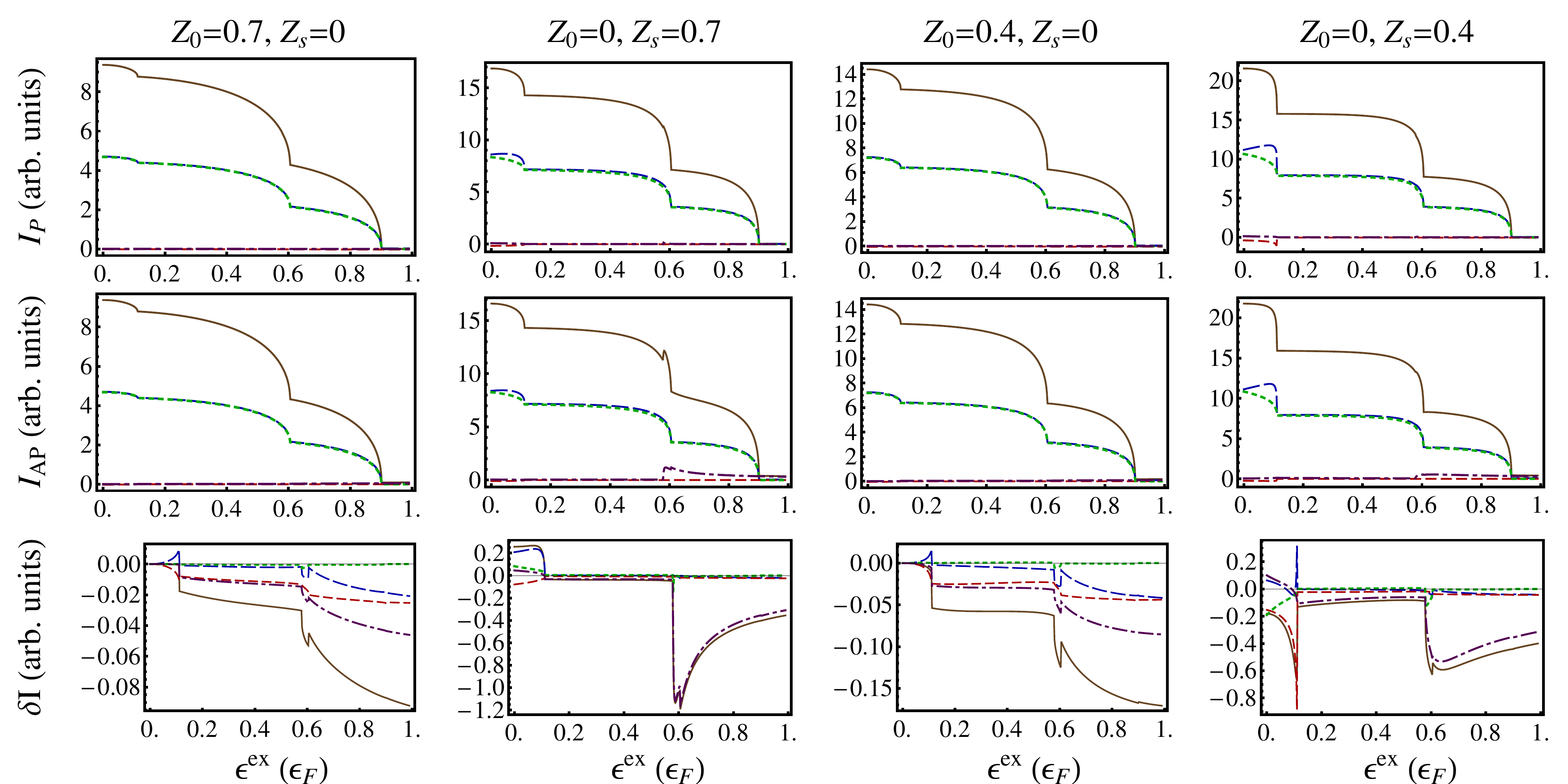}
  \caption{(Color online) Local normal, local Andreev,
  crossed normal, cross Andreev currents, and their combined effect (blue long-dashed,
  green dotted, red short-dashed, purple dot-dashed, and brown solid curves,
  respectively) vs exchange energy, $\epsilon^{\rm{ex}}$.
  From top to bottom rows presents the current of the P case,
  $I_{\rm{P}}$, AP case, $I_{\rm{AP}}$, and the difference,
  $\delta I=I_{\rm{P}}-I_{\rm{AP}}$, respectively.
  The first and second columns correspond to a spin-independent
  barrier and a purely SDIB of strength $0.7$, respectively,
  while the third and fourth correspond to strength $0.4$.
  Remarkably, the dot-dashed curves (contributed by CAR)
  at $\epsilon^{\rm{ex}}>0.6\epsilon_F$ in panels of the
  bottom row show much larger values for SDIB (non-zero $Z_s$) than those for the spin-independent barrier
  ($Z_s=0$), indicating a strong enhancement of CAR in the P case.
  The coincidence of two curves in some regions indicates the two local (crossed) currents
  almost equal to each other.
  }
        \label{fig:f05}
\end{figure*}

\subsection{Effects of ferromagnetism}

The effects of ferromagnetism on the charge transport emerge in
two ways. The first is that the presence of $\epsilon^{\rm{ex}}$
in the F leads results in a density imbalance in spin species and
thus alters both the number of channels and the momentum of
particles [see Eq.~(\ref{eqn:WV12})] responsible for carrying
charge current. The second is that $\epsilon^{\rm{ex}}$ in the
interface region constitutes SDIB parameters $Z$ (as discussed in
Sec.~\ref{sec:model}), which lead to different scattering phases
between majority and minority spins, and in turn to interference
in the scattering wave function, as we will see in subsequent
discussions and results.

Mathematically, this can be seen by noting in Eq.~(\ref{eqn:PCF})
that the current is a product of particle momentum $p$ associated
with the first effect, and scattering magnitudes $|a|$ and $|b|$
altered mainly by the second effect. The number of channels,
$M^j_{a/b}$ (discussed in Sec.~\ref{sec:BTK}), also reflects the
imbalance effect and plays a role when we add up the probability
current in Eq.~(\ref{eqn:Ipar}) and (\ref{eqn:Itot}). Both effects
contribute to a difference in behavior between the P and AP cases
solely due to the presence of coherent crossed transport (CNR and
CAR). This can be seen by noting that in the absence of these
processes, for instance, for large separation between F leads,
each lead can be treated independently and thus the P and AP cases
would show the same results. On the other hand, the CNR and CAR
processes distinguish the fact that in the P case the majority
spin species are the same in both leads but are different in the
AP case. We expect that the first effect is obscured in the low
imbalance regime where the majority and minority species are less
distinguishable, while the second effect is suppressed when the
interface barrier is nearly spin independent ($Z_s \sim 0$).

In this subsection we choose the set of parameters
$W_{1}=W_{2}=L=10k_F^{-1}$, $W_S=100k_F^{-1}$, and zero
temperature ($T=0$). Figure \ref{fig:f02} shows $\delta R$ as a
function of a purely SDIB ($Z_0=0$ and $Z_s$ varied, solid curve)
and a spin-independent case ($Z_s=0$ and $Z_0$ varied, dashed
curve) at a small imbalance of
$\epsilon^{\rm{ex}}=0.01\epsilon_F$. We see that a purely SDIB
causes an obvious variance of $\delta R$ in both positive and
negative values at an intermediate $Z_s$. At large $Z_s$ the
barrier is high enough such that most incidence is directly
reflected while at small $Z_s$ the phase shift between majority
and minority spins is close to zero, both of which make the
difference between P and AP cases negligible and hence give
$\delta R \sim 0$. For the spin-independent barrier, there is
almost no difference between P and AP, so $\delta R$ is always
flat and close to zero (still negative, reflecting the small
imbalance effect as discussed before). A complete dependence of
$\delta R$ on both $Z_0$ and $Z_s$ is shown in the contour plot in
Fig.~\ref{fig:f03}. We see that $\delta R$ is negative for most of
the parameter space and only becomes positive when $Z_s$
dominates.

Although the maximum magnitude in the positive region is one order
smaller than that in the negative one, there is a wide enough
parameter regime for $0.1\%<\delta R<1.5\%$, the same order
magnitude observed in the experiment of Ref.~\onlinecite{Colci12}.
The most negative value is centered around the region of line
$Z_0=Z_s$ where either majority or minority spins are subject to
an almost transparent interface. This is because in the AP case
different spin species in the two F leads see the transparent
barrier, so the CAR is greatly enhanced and hence reduces the
resistance compared to the P case. Notice that there is a small
positive region at $Z_s<Z_0$ with $0<\delta R<10^{-4}$. However,
for $Z_s=0$, $\delta R$ is always negative, indicating that the
SDIB is essential for positive $\delta R$. The contour plot of
Fig.~\ref{fig:f03} remains unchanged if $Z_s \rightarrow -Z_s$.

We now analyze the effect of density imbalance. Figure
\ref{fig:f04} shows comparison of $\delta R$ as a function of
$\epsilon^{\rm{ex}}$ for two purely SDIB conditions where $\delta
R$ is initially positive ($Z_s=0.7$, solid curve) and negative
($Z_s=0.4$, dashed) at zero exchange, as well as for  a
spin-independent condition ($Z_0=0.7$, dotted). First we see that
at large imbalance all curves drop to the highly negative region.
This is due to a combination of number of channels and the
momentum for the minor species monotonically decreasing with the
increase in $\epsilon^{\rm{ex}}$, as shown in
Eq.~(\ref{eqn:WV12}). This decrease reduces current carried by all
Andreev processes involving one majority and one minority spin;
CAR in AP case is not reduced as it involves two majority spins in
the two F leads. At large imbalance in the P case most incoming
current directly reflects and leaves small net current compared to
AP case. Therefore $\delta R$ drops and finally reaches its
maximum value of $-2$ [see Eq.~(\ref{eqn:DR})] at which
$I_{\rm{P}}=0$ (see inset).

The imbalance has a monotonic influence on the transport,
explaining the curve for the case of spin-independent barrier. For
an SDIB, however, $\delta R$ exhibits an overall decreasing trend
but a locally (e.g., in $0.5 < \epsilon^{\rm{ex}}/\epsilon_F
<0.7$) non-monotonic form  as a function of $\epsilon^{\rm{ex}}$.
We attribute this behavior to an interference effect of scattering
through the SDIB. Differing interference effects between the P and
AP cases can thus either enhance or compete with the imbalance
effect and dominate over it to make $\delta R$ largely negative or
even positive at low imbalance.

We further investigate the imbalance effect by looking at four
different contributions to the total current: the currents carried
by (1) the incoming electrons and LNR, (2) CNR, (3) LAR, and (4)
CAR [$I^{\rm{LN}}$, $I^{\rm{CN}}$, $I^{\rm{LA}}$, and
$I^{\rm{CA}}$, respectively, defined in Eq.~(\ref{eqn:Ipar5}) in
Appendix \ref{sec:I-V}]. Figure \ref{fig:f05} shows these
components and their combined effect as a function of
$\epsilon^{\rm{ex}}$ for P (top row) and AP (middle row) cases as
well as the difference between P and AP (bottom row) at various
interface conditions (corresponding to different columns). We
first see that the SDIB cases have more current (lower resistance)
than the spin-independent cases of the same barrier strength. For
all cases the two local currents are the dominant contributions to
the total current and are comparable in magnitude. Their
monotonically decreasing trends are consistent with an increase in
imbalance, except $I^{\rm{LN}}$ in the low imbalance regime
($\epsilon^{\rm{ex}}<0.11\epsilon_F$) in the case of a purely SDIB
of $Z_s=0.4$ (the rightmost column), which reflects a stronger
interference effect than the imbalance effect. The two crossed
currents have such small contributions compared to the local
current that they are almost indiscernible from zero in the
figure, except $I^{\rm{CA}}$ in the high imbalance regime in the
two SDIB cases, which indicates a great enhancement in long range
entanglement. The difference in current between P and AP cases,
$\delta I=I_{\rm{P}}-I_{\rm{AP}}$ [directly related to $\delta R$
via Eq.~(\ref{eqn:DR})], is one to two orders smaller than either
$I_{\rm{P}}$ or $I_{\rm{AP}}$ and is highly sensitive to the value
of the exchange field.

The two spin-independent cases have a similar behavior: At low
imbalance the compensation of dominant contributions of $\delta
I^{\rm{LN}}$ and $\delta I^{\rm{CN}}$ makes the total $\delta I$
close to zero (still negative), while at high imbalance $\delta
I^{\rm{CA}}$ has a large negative contribution and $\delta
I^{\rm{LA}}$ is suppressed, as supported by the imbalance effect.

The purely SDIB cases have various dominant contributions at low
imbalance. At $Z_s=0.7$ (the second column from left), the
dominant contribution is positive $\delta I^{\rm{LN}}$, making the
sum of the contributions positive, while at $Z_s=0.4$ (the
rightmost column) the strong negative $\delta I^{\rm{CN}}$ makes
the sum negative. These results can only be attributed to the
interference of scattering through an SDIB. At high imbalance,
$\delta I^{\rm{CA}}$ dominates as in the spin-independent case but
is much larger, due to the greater enhancement of $I^{\rm{CA}}$ in
the AP case. Notice that all the curves have kinks at the same
positions---this is due to the reduction in  the number of
scattering channels by one at each kink.

In brief, we analyzed the effects of spin imbalance and scattering
through an SDIB, both induced by ferromagnetism. The former
monotonically lowers $\delta R$ toward negative values and
dominates in the high imbalance regime, while the latter exhibits
its influence in positive or negative directions, where the trend
is revealed in the scattering magnitudes. A system with an SDIB
can sustain a significant large CAR current in the AP case,
reminiscent of a great enhancement of long range entanglement.

\begin{figure}[t]
\centering
   \includegraphics[width=6.5cm]{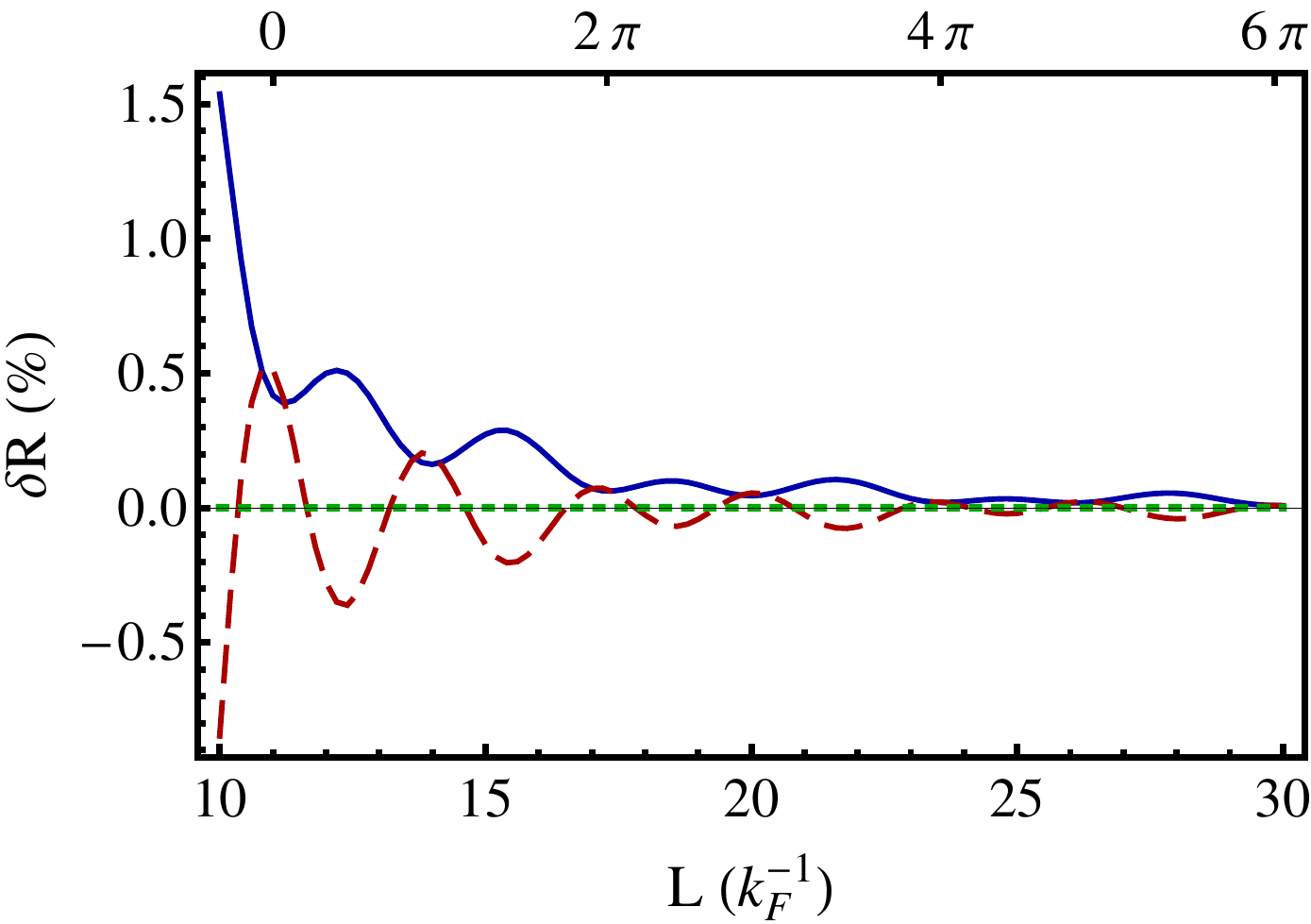}
  \caption{(Color online) Relative resistance, $\delta R$,
  vs the separation between two F leads, $L$,
  with fixed $W_{1(2)}$ ($W_1=W_2=10k_F^{-1}$)
  at the interface parameters $\{Z_0,Z_s\}=\{0,0.7\}$ (solid curve),
  $\{0,0.4\}$ (dashed), and $\{0.7,0\}$ (dotted),
   same as presented in Fig.~\ref{fig:f04}. Relative distance
   from a reference is marked on top of graph, showing the oscillatory
  period of $\pi$.}
        \label{fig:f06}
\end{figure}

\begin{figure}[t]
\centering \subfigure[]{\includegraphics[width=7cm]{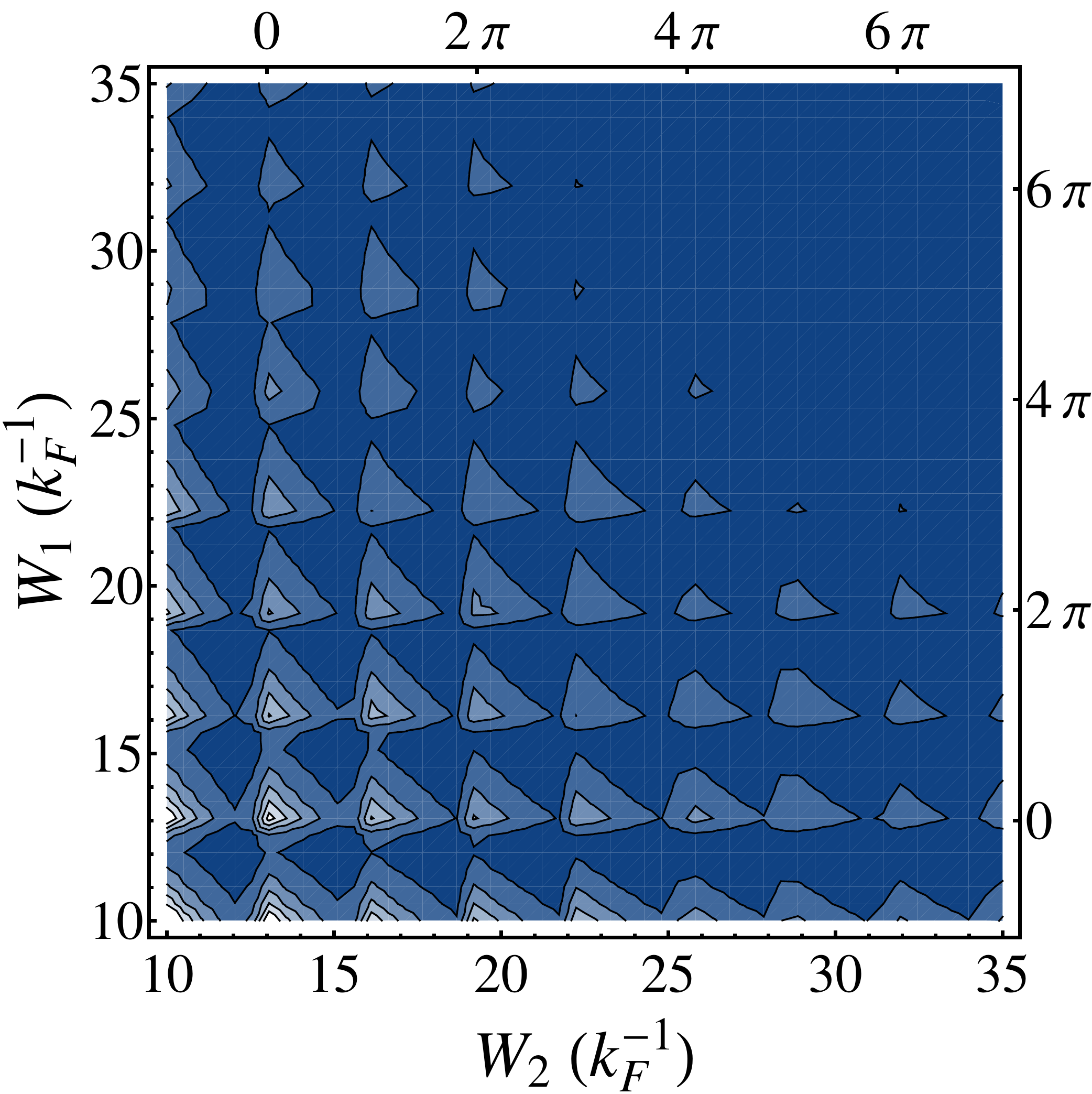}}
\subfigure[]{\includegraphics[width=7cm]{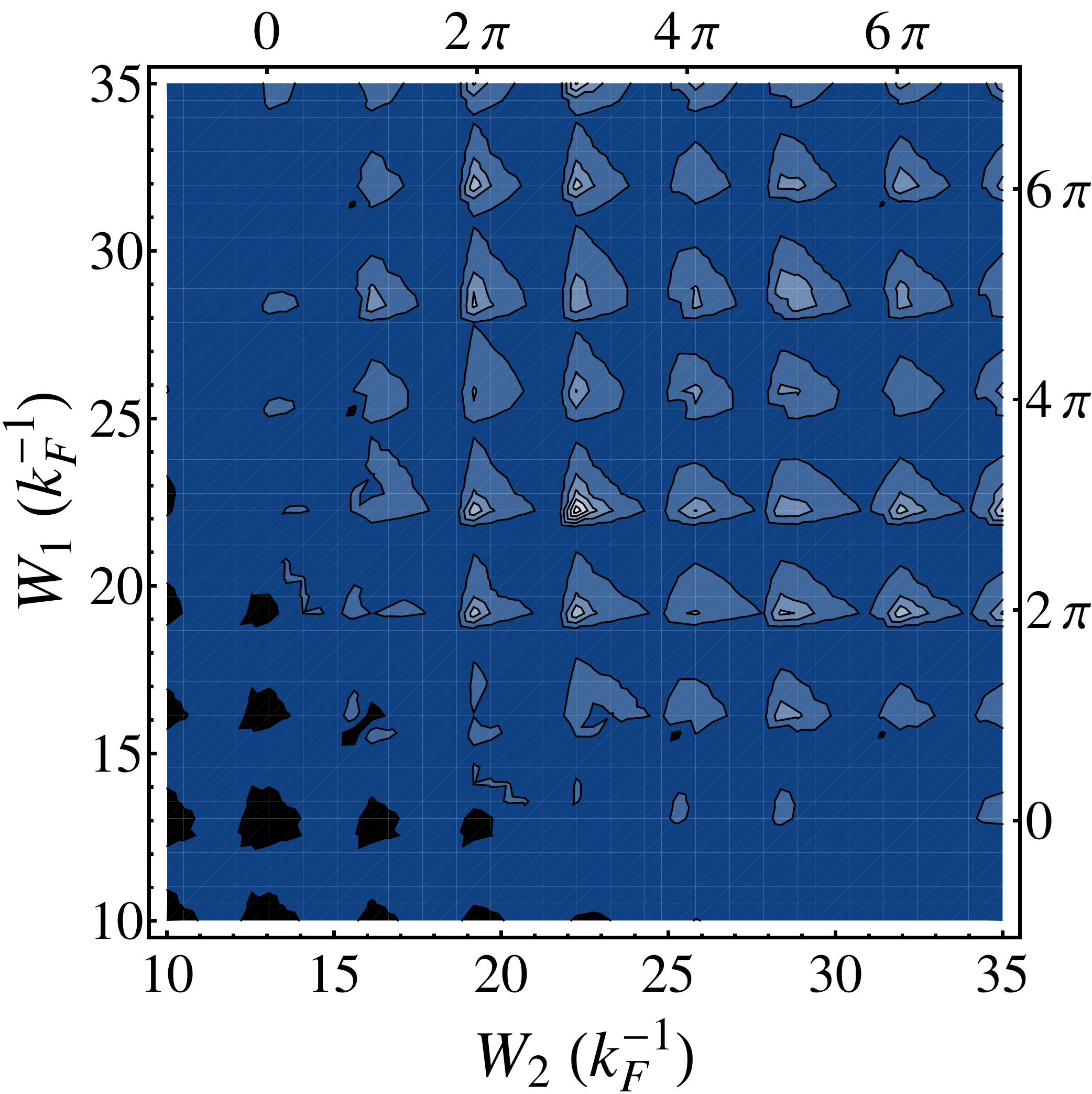}}
\caption{(Color online) Contour plot of $\delta R$ in the plane of
  $W_1$ and $W_2$. Data presented for a pure SDIB of $Z_s=0.7$ (a) and
  $0.4$ (b). The lowest contour is zero, below which the value
  is negative (corresponding to the black region).
  The magnitude difference between the contours is $0.2\%$. Relative distance
   from a reference is marked on top and right of graph, showing the oscillatory
  period of $\pi$.
   }
        \label{fig:f07}
\end{figure}

\begin{figure}[t]
\centering
   \includegraphics[width=7cm]{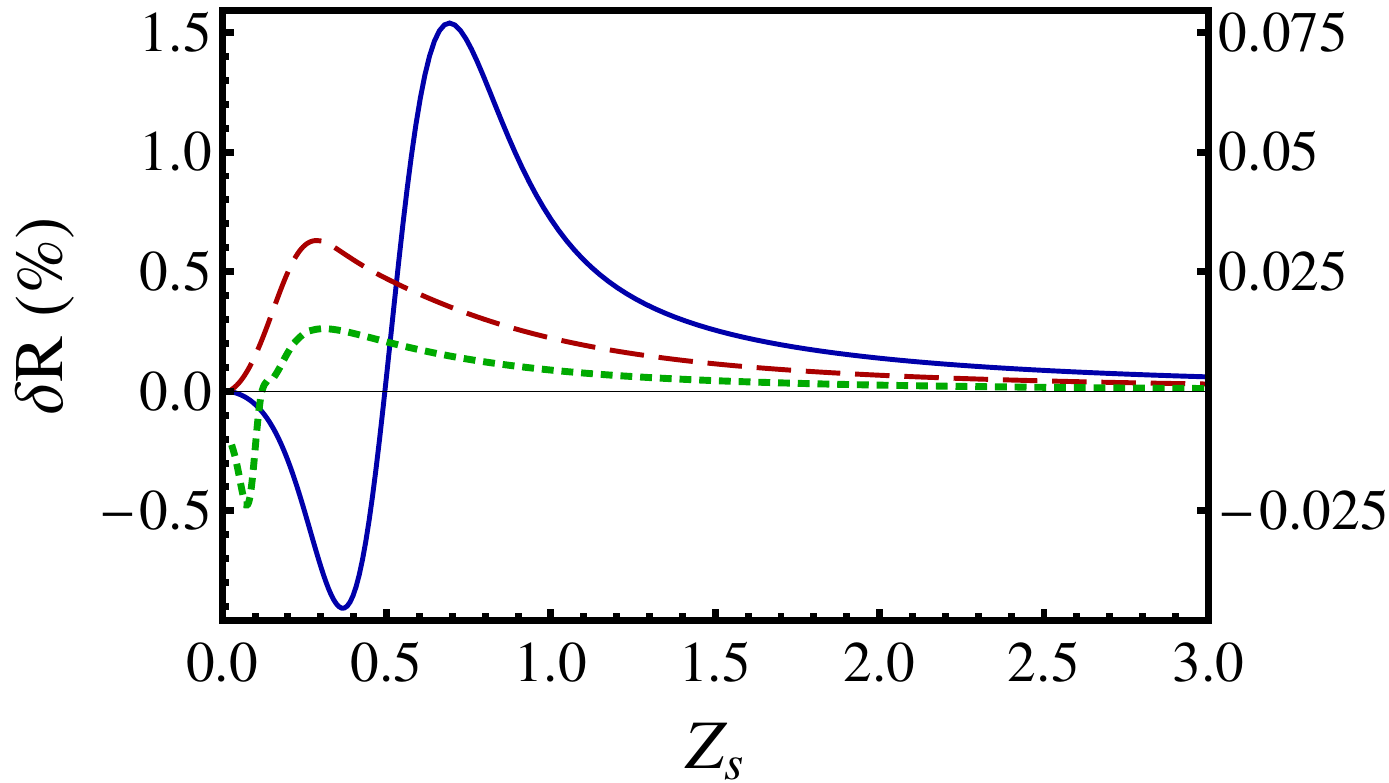}
  \caption{(Color online) Relative resistance $\delta R$
  vs pure SDIB $Z_s$. The solid curve (axis on the left
  of the graph) presents a purely SDIB system of
  $W_{1}=W_{2}=L=10k_F^{-1}$ and $W_S=100k_F^{-1}$
  (same as presented in Fig.~\ref{fig:f02}).
  The dashed and dotted curves (axis on the right
  of the graph) are for devices with sizes $10$ and $30$ times
  that for the solid curve, respectively. }
        \label{fig:f08}
\end{figure}

\subsection{Effects of geometry}

The signature of interference due to scattering between the two F
leads implies a strong dependence of the transport properties on
their geometry. We examine $\delta R$ as a function of the widths,
$W_1$ and $W_2$, and the separation, $L$, of the leads at a low
imbalance value of $\epsilon^{\rm{ex}}=0.01\epsilon_F$ where the
scattering effects dominate. Here we fix $W_S=100k_F^{-1}$. Figure
\ref{fig:f06} shows $\delta R$ vs $L$ at three barrier conditions,
$\{Z_0,Z_s\}=\{0,0.7\}$, $\{0,0.4\}$, and $\{0.7,0\}$, as
presented in Fig.~\ref{fig:f04}. For the two SDIB cases, the
curves modulate and exponentially decay to zero at large $L$. The
oscillation period is $\pi$ (in units of $k_F^{-1}$), which
results from a $2\pi$ modulation in the scattering amplitudes,
which in turn owe to the oscillatory behavior of matrix $\Omega$
in Eq.~(\ref{eqn:AM12b}). Since the quantity $\Lambda$ in
Eq.~(\ref{eqn:LAMBDA}) is a sinusoidal function in $W_{1/2/S}$ and
$L$, summing the product of $\Lambda$ in calculating $\Omega$
directly indicates a relation between the interference and the
geometry of the system. At large L, the scattering hardly shows a
difference between P and AP cases. The fast decay as a function of
$L$ is governed by a power-law factor in $L/k_F^{-1}$ due to the
interference rather than an exponential factor associated with the
Pippard coherence length ($\sim 500 k_F^{-1}$) (discussed in
Ref.~\onlinecite{Yamashita03}). For the spin-independent barrier,
the curve is flat and close to zero as expected. Notice that the
modulations in the current for P and AP cases still exists (as
discussed in Ref.~\onlinecite{Yamashita03}). However, they
modulate in-phase and  cancel out in the form of $\delta R$.

Figure \ref{fig:f07} shows contour plots of $\delta R$ in the
plane of $W_1$ and $W_2$ at $L=(W_1+W_2)/2$ for the pure SDIB
cases of $Z_s=0.7$ [panel (a)] and $0.4$ [panel (b)]. We see
$\delta R$ modulates with a period of $\pi$ in both $W_1$ and
$W_2$ directions, for the same reason as for variation in $L$. In
(a), $\delta R$ is always positive and becomes close to zero at
large $W_1$ and $W_2$. In (b), the value is mainly positive with a
slower decay as $W_1$ and $W_2$ go large. Both plots show a wide
parameter range for positive $\delta R$ in contrast to the small
negative value expected in the case of spin-independent barriers.

Finally we keep relative ratios between $W_1$, $W_2$, $W_S$, and
$L$ unchanged and enlarge the whole device (making it closer to
some of the realistic systems~\cite{Colci12}). Figure
\ref{fig:f08} shows $\delta R$ as a function of a pure SDIB
parameter $Z_s$ for three difference scales of the device: the one
of $W_{1}=W_{2}=L=10k_F^{-1}$ and $W_S=100k_F^{-1}$ (solid curve,
same as presented in Fig.~\ref{fig:f02}) as well as $10$ and $30$
times the case of the solid curve (dashed and dotted curves,
respectively). When the system becomes an order of magnitude
larger, the positive $\delta R$ region expands but its value is
more than one order smaller. Because larger sizes result in more
conducting channels and hence more interference among them, this
trend can smear out the difference between P and AP cases and
leave a small $\delta R$. To improve the modeling for a real
system that can still have $\delta R$ of a few percent, new
ingredients such as the dependence of $Z$ parameters on spatial
coordinate or on channels would be needed.

\begin{figure}[t]
\centering
   \includegraphics[width=6.5cm]{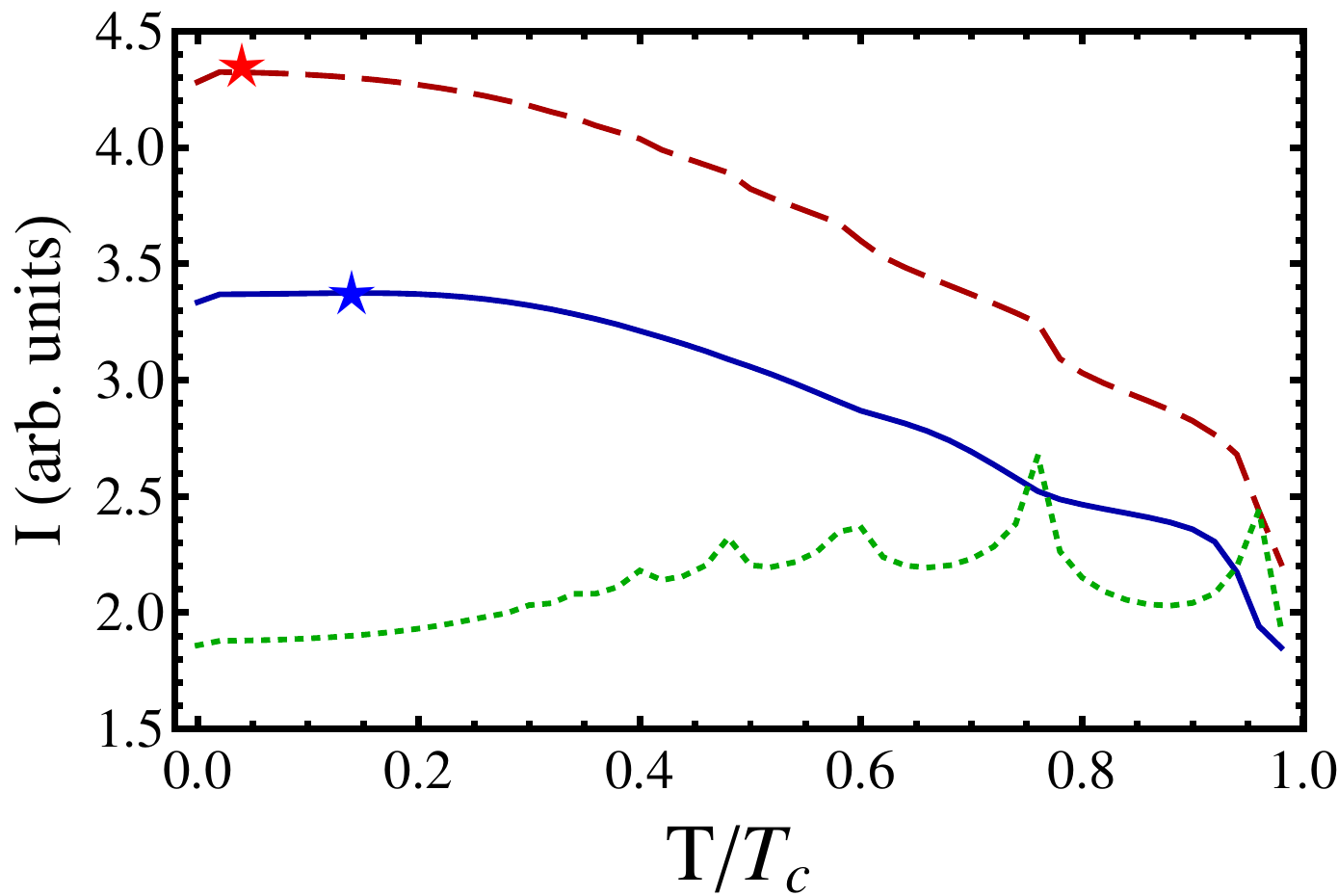}
  \caption{(Color online) Charge current $I$
  vs the temperature $T$ for the P case at the interface
  parameters $\{Z_0,Z_s\}=\{0,0.7\}$ (solid curve),
  $\{0,0.4\}$ (dashed), and $\{0.7,0\}$ (dotted),
  as presented in Fig.~\ref{fig:f04}. The star signs indicate
  position of the maximum point of the corresponding curves.}
        \label{fig:f09}
\end{figure}

\begin{figure}[t]
\centering
   \includegraphics[width=6.5cm]{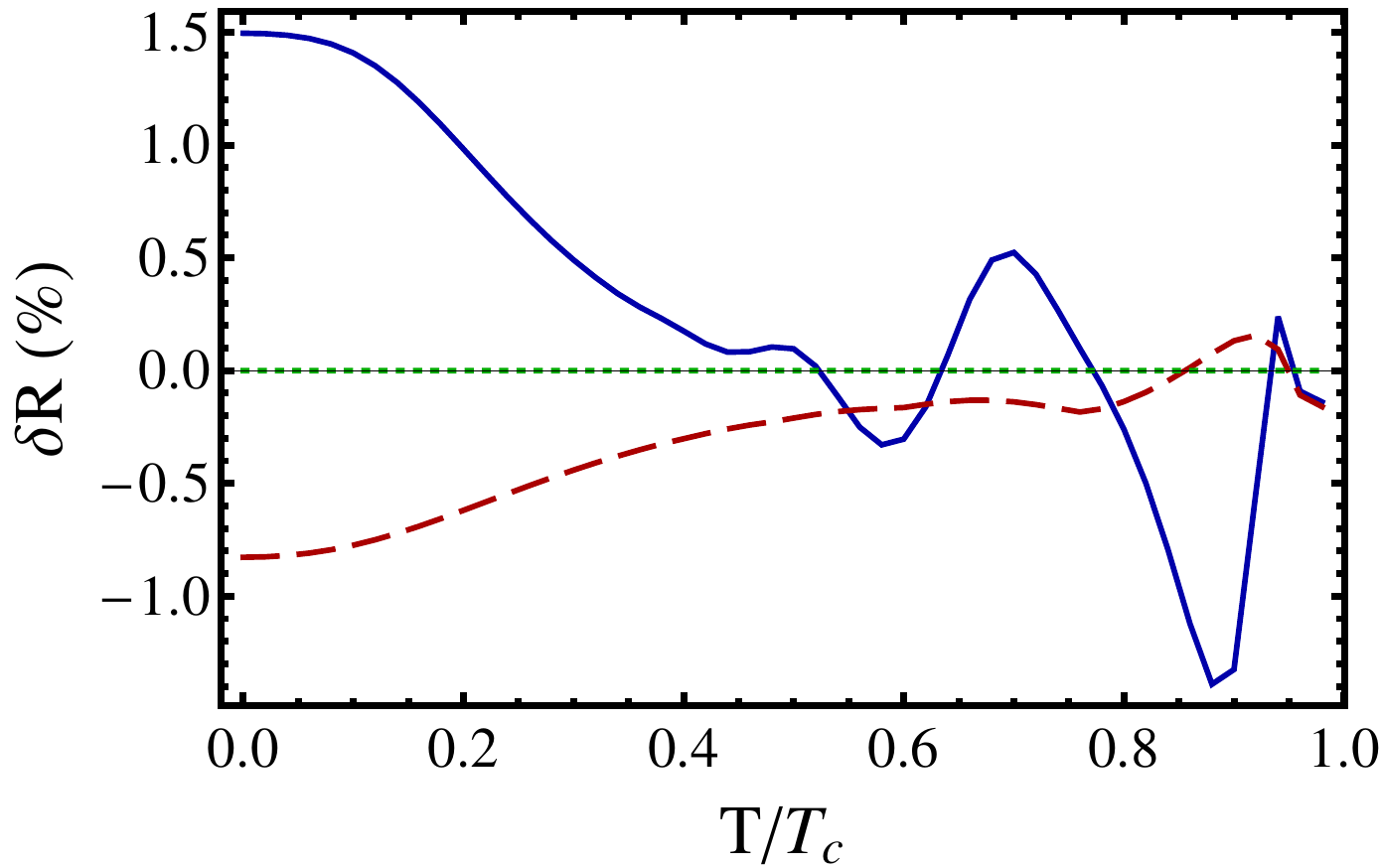}
  \caption{(Color online) Relative resistance $\delta R$
  vs the temperature $T$ at the interface
  parameters $\{Z_0,Z_s\}=\{0,0.7\}$ (solid curve),
  $\{0,0.4\}$ (dashed), and $\{0.7,0\}$ (dotted), convention
  as presented in Fig.~\ref{fig:f09}. Notice that the dotted curve almost
  coincide with the axis.}
        \label{fig:f10}
\end{figure}

\subsection{Effects of temperature}
Finally, we briefly study finite-temperature effects. Here, two
factors need to be incorporated: (1) quasi particle (hole)
transmissions (QPT and QHT, respectively) that arise from thermal
fluctuations and participate in charge transport as well as (2)
the decrease in the superconducting gap as a function of
temperature, $\Delta(T)=\Delta(0)\sqrt{1-T/T_c}$. Here, we focus
on the effect of the SDIB and thus choose the case of low
polarization for which the effects of SDIB are most prominent. We
plot the current in the P case, $I_{\rm{AP}}$, vs temperature,
$T$, for different barrier conditions in Fig. \ref{fig:f09}. All
cases show suppressed conductance for a significant temperature
regime close to $T_c$ due to the vanishing superconductivity. The
spin independent interface shows local peaks which we attribute to
thermal population of new channels. Spin dependence of the barrier
suppresses these peaks. We also mention that in accordance with
BTK expectations,~\cite{Blonder82} while the SDIB curves seem to
decrease as a function of temperatures, the maximum (star signs in
graph) occurs at non-zero but very low temperature, indicating a
reentrant effect.

To characterize the difference between P and AP cases, we plot
$\delta R$ vs $T$ for the same set of parameters in
Fig.~\ref{fig:f10}. We see that $\delta R$ can be smooth, widely
fluctuating, or even endowed with a sign change for different SDIB
conditions. The salient feature here is that in the
spin-independent case, due to the low polarization, $\delta R$ is
exceedingly small and constant on the scale shown. However the
SDIB cases show that temperature can alter the sign of $\delta R$
and what might have started as negative at low temperature can
reverse the trend. The lack of trend has also been found in
different samples in experiment.~\cite{Colci12}

\section{Conclusion}\label{sec:conclusion}

In this paper, we studied electronic transport physics in
superconductor--double ferromagnet (S-FF) junctions with
spin-dependent interface barriers (SDIB) using the
Blonder-Tinkham-Klapwijk (BTK) treatment. We proposed
spin-dependent $Z$ parameters modeling SDIB for the BTK
calculations and discussed its relation to microscopic physical
quantities at the interface. Our extended model is one of the
first incorporating SDIB for S-FF junctions and shows that SDIB
can increase crossed Andreev reflection (CAR) current. We found
that SDIB can cause interference in wave functions of CAR and
other scattering processes that can either collaborate or compete
with the imbalance effect in the F leads. This competition can
cause more current carried by F leads of parallel (P)
magnetization than by the antiparallel (AP) case (resulting in a
positive $\delta R$), which does not occur in the case of
spin-independent barriers. Our calculations showed that positive
$\delta R$ is stable in a sufficiently large parameter regime as a
function of $Z$ parameters, magnetization, geometry of the system,
and temperature. At low magnetization, the effects of SDIB are
responsible for the different transport between P and AP cases,
resulting in positive or negative $\delta R$ that is sensitive to
the $Z$ parameters. At high magnetization, $\delta R$ is always
negative due to the population imbalance in the F leads, and SDIB
contributes large additional enhancement of CAR in the AP case.
Variance in the geometry of the junction results in oscillations
in $\delta R$, which reveal the interference effects on the
transport physics.

Our studies suitably describe recent experiments of
Ref.~\onlinecite{Colci12} and can be used as a guide for further
investigation in such settings, including the possibility of
controlled tuning of S-F interface properties in the presence of
artificially grown SDIB barriers. Our model easily extends to a
wide class of multi terminal hybrid junctions. An important
ingredient to take into account in the future, particularly to
explain differing P versus AP behavior, is to incorporate induced
spin-triplet correlations together with the spin-flip
scatterers.~\cite{Volkov03,Bergeret05,Keizer06,Eschrig08,Linder10,Almog11}
This would be particularly relevant to currently intense
explorations of coherent properties and long-range entanglement in
superconductor heterostructures.

\section*{ACKNOWLEDGEMENT}
This work is supported by the DARPA-ARO Award No. W911NF-07-1-0464
(K.S.), by the University of Cincinnati (N.S.), and by NSF Grant
No. DMR-0906521 (S.V.). N.S. and S.V. acknowledge the hospitality
of the Aspen Center for Physics. We are grateful to Madalina Colci
and Dale J. Van Harlingen for motivating these studies and for
valuable input.

\appendix
\section{Semiclassical picture for SDIB}\label{sec:semiclassical}
In this section we use a semiclassical picture to model the
spin-dependent component of the SDIB parameter, $Z_s$, in
Eq.~(\ref{eqn:Zsig}). In the original BTK
treatment,~\cite{Blonder82} $Z$ is considered as a
phenomenological barrier parameter capable of characterizing
distinct properties of the interface, such as a metallic or
insulating one. In our extended model for the SDIB, the $Z$
parameters ought to reflect general concepts in the spin-related
transport, such as the spin-dependent interfacial phase
shift.~\cite{Cottet05} Based on this assumption, we derive $Z_s$
by relating the phase shifts between scattering through the
phenomenological $\delta$-function barrier and that through the
microscopic Zeeman potential at the interface, provided the
spin-independent component $Z=0$. Microscopically, an injected
spin passes through the barrier region and accumulates a phase
shift  $\phi$ in the interface region, compared with the case
without a barrier. In a semi-classical picture, the phase shift
due to the deviation of the exchange energy (compared with the
no-barrier case), $\delta {{\epsilon ^{{\rm{ex}}}}(x)}$, is
\begin{eqnarray}
{\phi } &=& \int\limits_0^{{t_0}} {\frac{{\delta {\epsilon
^{{\rm{ex}}}}(x(t))}}{\hbar }dt}  = \int\limits_0^\xi
\frac{{\delta\epsilon ^{{\rm{ex}}}}(x)}{\hbar {v_F}}dx \nonumber\\
&\equiv& \frac{{ \xi \langle {{\delta \epsilon
^{{\rm{ex}}}}}\rangle }}{{\hbar {v_F}}},\label{eqn:Phi}
\end{eqnarray}
where $t_0$ is the total time to pass the barrier, $v_F$ is the
Fermi velocity, ${\delta \epsilon ^{{\rm{ex}}}}(x(t))$ is the
local deviation of exchange energy when the particle passes the
position $x$ at time $t$, and $\langle {{\delta \epsilon
^{{\rm{ex}}}}}\rangle$ is its average over the width of the
interface, $\xi$. Here $x = v_F t$ given that the energy variance
is small compared to the Fermi energy.

Now we compare the phase shifts with those of a scattering
problem having the $\delta$-function potential,
\begin{eqnarray}
- \frac{{{\hbar ^2}}}{{2m}}\partial_x^2 + \hbar {v_F}{Z_s }\delta
(x).
\end{eqnarray}
By solving the Schr\"odinger equation for an incoming wave around
the Fermi surface, we obtain $Z_s$ as a function of the phase
shift of the transmitted wave,
\begin{eqnarray}
{Z_s } =  - \tan {\phi}.\label{eqn:ZPhi}
\end{eqnarray}
Substituting Eq.~(\ref{eqn:ZPhi}) into Eq.~(\ref{eqn:Phi}), we
derive the aimed relation of Eq.~(\ref{eqn:Zp}). The tangent
function implies that ${Z_s}$ in non-monotonic in and sensitive to
the exchange energy and microscopic length. Notice that the
majority and minority spins have opposite exchange energy which
leads to different signs for the spin-dependent $Z$ parameter. In
this paper we choose $\pm Z_s$ for majority (minority) spins, as
shown in Eq.~(\ref{eqn:Zsig}). Notice that our focus here is one
of the scattering-related causes for an SDIB. We perceive that
other effective treatments, such as first-principles
methods~\cite{Stiles96,Schep97,Xia02}, can be applied for further
analysis.

\section{$I$-$V$ relation}\label{sec:I-V}
In this section we detail the calculations for current-voltage
relations in the SFF junction following the standard BTK
formalism~\cite{Blonder82,Yamashita03} and identify the current
contributions for each scattering process. The total current
contributed by an incoming wave characterized by indices
 $\{{\tau \sigma jn}\}$ is given by
\begin{eqnarray}
{I_{\tau \sigma jn}} = A \tau \int\limits_0^\infty  {dE[{f_ \to
}(E) - {f_ \leftarrow }(E)]},\label{eqn:Ipar1}
\end{eqnarray}
where $A$ is a constant associated with density of states, Fermi
velocity, and an effective cross-sectional area and
$f_{\to(\leftarrow)}$ denotes the incoming (outgoing) populations
\begin{eqnarray}
{f_ \to }(E) &=& {f_0}(E - \tau eV),\label{eqn:Ipar2}\\
{f_ \leftarrow }(E) &=& \sum\limits_{\{ \tau ',\sigma '\}  =  \pm
\{ \tau,\sigma\} } {\sum\limits_{j'} {\sum\limits_{n'} {\left|
{\tilde J_{\tau \sigma jn}^{\tau '\sigma 'j'n'}}
\right|{f_0}(E - \tau ' eV)  } } }\nonumber\\
&{}&  + \sum\limits_{\tau ' =  \pm } {\sum\limits_{n'} {\left|
{\tilde J_{\tau \sigma jn}^{{\rm{S}};\tau 'n'}} \right|{f_0}(E)}
}\label{eqn:Ipar3}.
\end{eqnarray}
Here we replace the distribution function for holes in the F
regions as $1 - {f_0}( - E - eV) = {f_0}(E + eV)$. By substituting
Eqs.~(\ref{eqn:Ipar2}) and (\ref{eqn:Ipar3}) into
Eq.~(\ref{eqn:Ipar1}) as well as applying a conservation relation
for the probability currents,
\begin{eqnarray}
\sum\limits_{\{ \tau ',\sigma '\}  =  \pm \{ \tau ,\sigma \} }
{\sum\limits_{j'} {\sum\limits_{n'} {\left| {\tilde J_{\tau \sigma
jn}^{\tau '\sigma 'j'n'}} \right|} } }  + \sum\limits_{\tau ' =
\pm } {\sum\limits_{n'} {\left| {\tilde J_{\tau \sigma
nj}^{{\rm{S;}}\tau 'n'}} \right|} }  = 1\nonumber
\end{eqnarray}
we eliminate the contributions from quasi particle transmissions
and hence derive the $I$-$V$ relation of Eq.~(\ref{eqn:Ipar}), as
a function of only processes in F regions. By summing over the
incoming degrees of freedom, we obtain the total current as in
Eq.~(\ref{eqn:Itot}), or simplified at zero temperature as
\begin{eqnarray}
&&I(V, T=0)\mathop  = A\sum\limits_{\sigma  = \pm
} {\sum\limits_n {\sum\limits_{j = 1,2} {\int\limits_0^{eV} {dE} }}}\nonumber\\
&&\times \bigg[ {1 + \sum\limits_{n'} {\sum\limits_{j' = 1}^2
{\left( {\left| {\tilde J_{ - \sigma nj}^{{\rm{ + }}\bar \sigma
n'j'}} \right| - \left| {\tilde J_{ + \sigma nj}^{ + \sigma n'j'}}
\right|} \right)} } } \bigg]. \label{eqn:Ipar4}
\end{eqnarray}
From Eq.~(\ref{eqn:Ipar4}) we identify four components of the
total current in the subgap regime below: the local normal current
${I^{{\rm{LN}}}}$ contributed by the incoming wave and LNR, as
well as the crossed normal current ${I^{{\rm{CN}}}}$, the local
Andreev current ${I^{{\rm{LA}}}}$, and the crossed Andreev current
${I^{{\rm{CA}}}}$, contributed by CNR, LAR, and CAR, respectively:
\begin{eqnarray}
{I^{{\rm{LN}}}} &=& A\sum\limits_{\sigma  =  \pm }
{\sum\limits_{n,n'} {\sum\limits_{j = 1,2} {\int\limits_0^{eV}
{dE} } } } \left( {1 - \left| {\tilde J_{ + \sigma nj}^{ + \sigma
n'j}} \right|} \right)\nonumber\\
{I^{{\rm{CN}}}} &=& A\sum\limits_{\sigma  =  \pm }
{\sum\limits_{n,n'} {\sum\limits_{j = 1,2} {\int\limits_0^{eV}
{dE} } } } \left( { - \left| {\tilde J_{ + \sigma nj}^{ + \sigma
n'\bar j}} \right|} \right)\nonumber\\
{I^{{\rm{LA(CA)}}}} &=& A\sum\limits_{\sigma  =  \pm }
{\sum\limits_{n,n'} {\sum\limits_{j = 1,2} {\int\limits_0^{eV}
{dE} } } } \left| {\tilde J_{ - \sigma nj}^{ + \bar \sigma
n'j(\bar j)}} \right|.\label{eqn:Ipar5}
\end{eqnarray}


\begin{thebibliography}{99}
%review articles and scattering model for SN junctions
\bibitem{Blonder82} G. E. Blonder, M. Tinkham, and T. M. Klapwijk, Phys. Rev. B \textbf{25}, 4515 (1982).
\bibitem{Beenakker97}C. W. J. Beenakker, Rev. Mod. Phys. \textbf{69}, 731 (1997).
\bibitem{Lambert98}C. Lambert and R. Raimondi, J. Phys.: Condens. Matter \textbf{10}, 901 (1998).
\bibitem{Eschrig09}M. Eschrig, Phys. Rev. B \textbf{80}, 134511 (2009).

%multi terminal devices and CAR
\bibitem{Byers95}J. M. Byers and M. E. Flatt\'e, Phys. Rev. Lett. \textbf{74}, 306 (1995).
\bibitem{Takahashi99}S. Takahashi, H. Imamura, and S. Maekawa, Phys. Rev. Lett. 82, 3911 (1999).
\bibitem{Deutscher00}G. Deutscher and D. Feinberg, Appl. Phys. Lett. \textbf{76}, 487 (2000).
\bibitem{Recher01}P. Recher, E. V. Sukhorukov, and D. Loss, Phys. Rev. B \textbf{63}, 165314 (2001).
\bibitem{Falci01}G. Falci, D. Feinberg, and F. W. J. Hekking, Europhys. Lett. \textbf{54}, 255 (2001).
\bibitem{Melin02}R. M\'elin and D. Feinberg, Eur. Phys. J. B \textbf{26}, 101 (2002).
\bibitem{Melin03}R. M\'elin and S. Peysson, Phys. Rev. B \textbf{68}, 174515 (2003)
\bibitem{Dong03}Z. C. Dong, R. Shen, Z. M. Zheng, D. Y. Xing, and Z. D. Wang, Phys. Rev. B \textbf{67}, 134515 (2003).
\bibitem{Russo05}S. Russo, M. Kroug, T. M. Klapwijk, and A. F. Morpurgo, Phys. Rev. Lett. \textbf{95}, 027002 (2005).
\bibitem{Morten06}J. P. Morten, A. Brataas, and W. Belzig, Phys. Rev. B \textbf{74}, 214510 (2006).
\bibitem{Metalidis10}G. Metalidis, M. Eschrig, R. Grein, and G. Sch\''on, Phys. Rev. B \textbf{82}, 180503 (2010).


%quantum computing
\bibitem{Burkard07}G. Burkard, J. Phys. Condens. Matter \textbf{19}, 233202 (2007).

%\proximity effect of SF junctions
\bibitem{Buzdin05}A. I. Buzdin, Rev. Mod. Phys. \textbf{77}, 935 (2005).
\bibitem{Bergeret05}F. S. Bergeret, A. F. Volkov, and K. B. Efetov, Rev. Mod. Phys. \textbf{77}, 1321 (2005).
\bibitem{Cottet05}A. Cottet and W. Belzig, Phys. Rev. B \textbf{72}, 180503(R) (2005).
\bibitem{Linder09}J. Linder, T. Yokoyama, A. Sudbo, and M. Eschrig, Phys. Rev. Lett. \textbf{102}, 107008 (2009).

%\spin-dependent interfacial scattering
\bibitem{Fogelstrom00} M. Fogelstr\"om, Phys. Rev. B \textbf{62}, 11812 (2000).
\bibitem{Barash02} Y. S. Barash and I. V. Bobkova, Phys. Rev. B \textbf{65}, 144502 (2002).
\bibitem{Kopu04} J. Kopu, M. Eschrig, J. C. Cuevas, and M. Fogelstro\"m, Phys. Rev. B \textbf{69}, 094501 (2004).
\bibitem{Faure06}M. Faur\'e, A. I. Buzdin, A. A. Golubov, and M. Yu. Kupriyanov Phys. Rev. B 73, 064505 (2006).

%\Andreev reflection in SF
\bibitem{deJong95}M. J. M. de Jong and C. W. J. Beenakker, Phys. Rev. Lett. \textbf{74}, 1657 (1995).
\bibitem{Giroud98}M. Giroud, H. Courtois, K. Hasselbach, D. Mailly, and B. Pannetier, Phys. Rev. B, \textbf{58}, R11872 (1998)
\bibitem{Zutic00}I. Zuti\`c and O. T. Valls, Phys. Rev. B \textbf{61}, 1555 (2000).
\bibitem{Strijkers01}G. J. Strijkers, Y. Ji, F. Y. Yang, C. L. Chien, and J. M. Byers, Phys. Rev. B \textbf{63}, 104510 (2001).
\bibitem{Kupferschmidt11} J. N. Kupferschmidt and P. W. Brouwer, Phys. Rev. B \textbf{83}, 014512 (2011).


%\Andreev reflection
\bibitem{Andreev64}A. F. Andreev, Sov. Phys. JETP \textbf{19}, 1228 (1964).

%\pi junction
\bibitem{Ryazanov01}V. V. Ryazanov, V. A. Oboznov, A. Yu. Rusanov, A. V. Veretennikov, A. A. Golubov, and J. Aarts, Phys. Rev. Lett. \textbf{86}, 2427 (2001).
\bibitem{Kontos02}T. Kontos, M. Aprili, J. Lesueur, F. Genet, B. Stephanidis, and R. Boursier, Phys. Rev. Lett. \textbf{89} 137007 (2002).
\bibitem{Robinson07}J. W. A. Robinson, S. Piano, G. Burnell, C. Bell, and M. G. Blamire, Phys. Rev. B \textbf{76}, 094522 (2007).
\bibitem{Kastening09} B. Kastening, D.K. Morr, L. Alff, and K. Bennemann, Phys. Rev. B \textbf{79}, 144508 (2009).


%SFFS experiments
\bibitem{Beckmann04}D. Beckmann, H. B. Weber, and H. v. L\"ohneysen , Phys. Rev. Lett. \textbf{93}, 197003 (2004).
\bibitem{Luo09}P. S. Luo, T. Crozes, B. Gilles, S. Rajauria, B. Pannetier, and H. Courtois, Phys. Rev. B \textbf{79}, 140508(R) (2009).
\bibitem{Colci10}M. Colci, Ph.D. thesis, U. of Illinois at Urbana-Champaign (2010).
\bibitem{Colci12}M. Colci, K. Sun, N. Shah, S. Vishveshwara, D. J. Van Harlingen, Phys. Rev. B \textbf{85}, 180512(R) (2012).


%SFF-BTK
\bibitem{Yamashita03} T. Yamashita, S. Takahashi, and S. Maekawa, Phys. Rev. B \textbf{68}, 174504 (2003).

%layer system
\bibitem{Aarts97}J. Aarts, J. M. E. Geers, E. Bruck, A. A. Golubov, R. Coehoorn, Phys. Rev. B \textbf{56}, 2779 (1997).
\bibitem{Gu02}J. Y. Gu, C.-Y. You, J. S. Jiang, J. Pearson, Y. B. Bazaliy, and S. D. Bader, Phys. Rev. Lett. \textbf{89}, 267001 (2002).
\bibitem{Giazotto06}F. Giazotto, F. Taddei, F. Beltram, and R. Fazio, Phys. Rev. Lett. \textbf{97}, 087001
(2006).

%nanowire system
\bibitem{Tian05}M. Tian, N. Kumar, S. Xu, J. Wang, J. S. Kurtz, and M. H. W. Chan, Phys. Rev. Lett. \textbf{95}, 076802 (2005).
\bibitem{Altomare06}F. Altomare, A. M. Chang, M. R. Melloch, Y. Hong, and C. W. Tu, Phys. Rev. Lett. \textbf{97}, 017001 (2006).

%spin-dependent phase shift
\bibitem{Tokuyasu88} T. Tokuyasu, J. A. Sauls, and D. Rainer, Phys. Rev. B \textbf{38}, 8823 (1988).

%current-correlation in SFF
\bibitem{Taddei02}F. Taddei and R. Fazio, Phys. Rev. B \textbf{65}, 134522 (2002).

%superconducting coherence length
\bibitem{Tinkham96}M. Tinkham, \emph{Introduction to Superconductivity} (McGraw-Hill,New York, 1996).
\bibitem{Leggett06}A. J. Leggett, \emph{Quantum Liquids} (Oxford University Press, Oxford, 2006).

%p-wave
\bibitem{Volkov03}A. F. Volkov, F. S. Bergeret, and K. B. Efetov, Phys. Rev. Lett. 90, 117006 (2003).
\bibitem{Keizer06}R. S. Keizer, S. T. B. Goennenwein, T. M. Klapwijk, G. Miao, G. Xiao and A. Gupta, Nature \textbf{439}, 825 (2006).
\bibitem{Eschrig08}M. Eschrig and T. Lofwander, Nature Physics \textbf{4}, 138 (2008).
\bibitem{Linder10}J. Linder, M. Cuoco, and A. Sudbo, Phys. Rev. B \textbf{81}, 174526 (2010).
\bibitem{Almog11}B. Almog, S. Hacohen-Gourgy, A. Tsukernik, and G. Deutscher, Phys. Rev. B \textbf{84}, 054514 (2011).

%details in SDIB
\bibitem{Stiles96}M. D. Stiles, J. Appl. Phys. \textbf{79}, 5805 (1996).
\bibitem{Schep97}K. M. Schep, J. B. A. N. van Hoof, P. J. Kelly, G. E. W. Bauer, and J. E. Inglesfield, Phys. Rev. B \textbf{56}, 10805 (1997).
\bibitem{Xia02}K. Xia, P. J. Kelly, G. E. W. Bauer, and I. Turek, Phys. Rev. Lett. \textbf{89}, 166603 (2002).

\end{thebibliography}
\end{document}